\documentclass[12pt]{article} 
\usepackage{amsmath}
\usepackage{amssymb}
\usepackage{amsfonts}
\usepackage{multirow}
\usepackage{natbib}
\usepackage{amsthm}
\usepackage{graphicx}
\usepackage{booktabs}
\usepackage[tight]{subfigure}
\usepackage{caption}
\usepackage{url}
\usepackage{setspace}
\usepackage{mathrsfs}
\usepackage[svgnames,usenames,dvipsnames]{xcolor}
\usepackage{mathtools}
\usepackage{graphicx}
\usepackage[margin=1in]{geometry}
\usepackage{float}
\usepackage{listings}
\usepackage{rotating}
\usepackage{subfigure}
\usepackage{xr}

\newcommand{\be}{\begin{equation}}
\newcommand{\ee}{\end{equation}}
\newcommand{\bea}{\begin{eqnarray}}
\newcommand{\eea}{\end{eqnarray}}
\newcommand{\mbf}[1]{\mathbf{#1}}
\newcommand{\mbs}[1]{\boldsymbol{#1}}

\newcommand{\by}{\mbf{y}}

\newcommand{\bd}{\mbf{d}}

\newcommand{\bu}{\mbf{u}}

\newcommand{\bv}{\mbf{v}}
\newcommand{\bx}{\mbf{x}}

\newcommand{\bA}{\mbf{A}}

\newcommand{\bD}{\mbf{D}}

\newcommand{\bQ}{\mbf{Q}}
\newcommand{\bV}{\mbf{V}}
\newcommand{\bX}{\mbf{X}}

\newcommand{\bW}{\mbf{W}}

\newcommand{\bP}{\mbf{P}}

\newcommand{\bmu}{\mbs{\mu}}
\newcommand{\bbeta}{\mbs{\beta}}

\newcommand{\bxi}{\mbs{\xi}}

\newcommand{\blambda}{\mbs{\lambda}}
\newcommand{\bomega}{\mbs{\omega}}
\newcommand{\bLambda}{\mbs{\Lambda}}

\newcommand{\bet}{{\mbs{\eta}}}

\newcommand{\bI}{\mbf{I}}
\newcommand{\bH}{\mbf{H}}

\newcommand{\ben}{\begin{equation*}}
\newcommand{\een}{\end{equation*}}
\newcommand{\bean}{\begin{eqnarray*}}
\newcommand{\eean}{\end{eqnarray*}}
\newcommand{\bsm}{\begin{smallmatrix}}
\newcommand{\esm}{\end{smallmatrix}}
\newcommand{\bmat}{\begin{matrix}}
\newcommand{\emat}{\end{matrix}}

\newcommand{\given}{\,|\,}
\newcommand{\T}{{\scriptscriptstyle \mathrm{T}}}

\newcommand{\bzero}{\mbs{0}}

\newcommand{\bSigma}{\mbs{\Sigma}}

\newcommand\indsim{\stackrel{\mathclap{ind}}{\sim}}

\newcommand{\bvarepsilon}{\mbs{\varepsilon}}

\begin{document}

   \begin{center}
       \vspace*{1cm}
       \large
	    \textbf{Leveraging national forest inventory data to estimate forest carbon density status and trends for small areas}\\
        \normalsize
          \vspace{5mm}
	    Elliot S. Shannon\textsuperscript{1,2,*}, Andrew O. Finley\textsuperscript{1,2}, Paul B. May\textsuperscript{3}, Grant M. Domke\textsuperscript{4}, \\Hans-Erik Andersen\textsuperscript{5}, George C. Gaines, III\textsuperscript{6}, Arne Nothdurft\textsuperscript{7}, Sudipto Banerjee\textsuperscript{8}
	   \\
        \vspace{5mm}
   \end{center}
   {\small 
   \begin{enumerate}
       \item Department of Forestry, Michigan State University, East Lansing, MI, USA.
       \item Department of Statistics and Probability, Michigan State University, East Lansing, MI, USA.
        \item Department of Mathematics, South Dakota School of Mines and Technology, Rapid City, SD, USA.
        \item USDA Forest Service, Northern Research Station, St. Paul, MN, USA. 
        \item USDA Forest Service, Pacific Northwest Research Station, Seattle, WA, USA.
        \item USDA Forest Service, Rocky Mountain Research Station, Missoula, MT, USA.
        \item Department of Forest and Soil Sciences, University of Natural Resources and Life Sciences, Vienna, Austria
        \item Department of Biostatistics, University of California, Los Angeles, Los Angeles, CA, USA.
       \end{enumerate}
       }
       \noindent $^\ast$ Corresponding Author: Elliot S. Shannon (shann125@msu.edu), 480 Wilson Rd, East Lansing, MI 48824, United States.

\clearpage

\section*{Abstract}

National forest inventory (NFI) data are often costly to collect, which inhibits efforts to estimate parameters of interest for small spatial, temporal, or biophysical domains. Traditionally, design-based estimators are used to estimate status of forest parameters of interest, but are unreliable for small areas where data are sparse. Additionally, design-based estimates constructed directly from the survey data are often unavailable when sample sizes are especially small. Traditional model-based small area estimation approaches, such as the Fay-Herriot (FH) model, rely on these direct estimates for inference; hence, missing direct estimates preclude the use of such approaches. Here, we detail a Bayesian spatio-temporal small area estimation model that efficiently leverages sparse NFI data to estimate status and trends for forest parameters. The proposed model bypasses the use of direct estimates and instead uses plot-level NFI measurements along with auxiliary data including remotely sensed tree canopy cover. We produce forest carbon estimates from the United States NFI over 14 years across the contiguous US (CONUS) and conduct a simulation study to assess our proposed model's accuracy, precision, and bias, compared to that of a design-based estimator. The proposed model provides improved precision and accuracy over traditional estimation methods, and provides useful insights into county-level forest carbon dynamics across the CONUS. 

\section{Introduction}\label{sec:introduction}

National Forest Inventory (NFI) programs provide critical information on forest health, sustainable management, and ecosystem change \citep{Wurtzebach_2019}. Increasingly, NFI data users require higher spatial and temporal resolution forest status and change parameter estimates than can be provided by the traditional design-based estimation methods used by NFI programs \citep{kohl2006sampling, breidenbach2012small, prisley2021}. Given the high costs associated with data acquisition, NFI measurements are typically collected over a sparse network of inventory plots, limiting reliable design-based estimates to large spatial and temporal scales \citep{bechtold_patterson_2005, tomppo_2010, WestfallFIA_2022}.

In response, small area estimation (SAE) methods have gained attention for estimating forest parameters in data-sparse settings \citep{schroeder2014improving, lister2020use, hou2021updating, coulston2021enhancing, finley2024, Shannon_2025}. ``Small areas'' are spatial, temporal, or biophysical extents with insufficient plot measurements for reliable direct estimates. Contemporary SAE methods employ statistical models to relate forest response variables to auxiliary data, improving accuracy and precision over design-based approaches when strong relationships exist between response variables and auxiliary information. Within a regression framework, auxiliary information enters as covariates, while residual spatial and temporal dependence is captured via structured random effects \citep{rao2015small}.

The Fay-Herriot (FH) model \citep{fay_herriot_1979} is widely used in SAE applications for NFI data \citep{VERPLANCK2018287, tomesgen_2021, cao_et_al_2022, Stanke_2022, Georgakis_2024}. A key reason for its popularity is that it is fit to small area direct estimates and therefore does not require exact NFI plot locations. This feature is particularly important because exact plot locations are not publicly available for most NFIs, as a measure to protect private forest land information and preserve the ecological integrity of the plots. In the FH model, the response variable is the direct estimate mean and viewed as a noisy realization of the true but unobserved (i.e., latent) super-population mean with a portion of this noise accounted for by the direct estimate's sampling variance. However, a major limitation to using the FH model is that direct estimates are often missing when sample sizes are too small or measurements are homogeneous \citep{Shannon_2025}. When all plot measurements in a small area are identical, the direct estimate variance is zero, rendering the estimate unusable in FH model specifications.

To address this limitation, we propose a Bayesian spatio-temporal SAE model that directly uses NFI plot-level measurements, bypassing the reliance on direct estimates. Although exact plot locations are often unavailable due to privacy restrictions, plot measurements can typically be assigned to small areas \citep{betchold_patterson_2015}. Our proposed model incorporates plot-level NFI data along with auxiliary covariates, such as remotely sensed tree canopy cover, to estimate forest parameters. This model accommodates spatially and temporally varying regression coefficients and integrates dynamic spatio-temporal effects via a random intercept term. Operating within a Bayesian framework, we quantify uncertainty through posterior distributions, employing efficient Markov chain Monte Carlo (MCMC) sampling techniques.

This study is motivated by the need for small area estimates of live forest carbon density (LFCD), which represents the carbon stored in live tree tissues. Accurate LFCD estimates are essential for understanding forests' roles in the global carbon cycle \citep{HURTEAU2021561} and are mandated by national and international reporting requirements \citep{UNFCCC, IPCC}.

The remainder of this paper is organized as follows: Section \ref{sec:methods} details the NFI data from the United States Forest Service Forest Inventory and Analysis (FIA) program, describes the proposed model, and outlines a simulation study. Section \ref{sec:results} presents results from the FIA data analysis and simulation study, assessing model accuracy and precision. Finally, Section \ref{sec:Discussion} provides a discussion, and Section \ref{sec:Conclusion} offers recommendations for practitioners.

\section{Methods}\label{sec:methods}

\subsection{Data}\label{sec:data}
\begin{figure}[ht!]
    \centering
    \includegraphics[trim={0.5cm 0cm 0.5cm 0.2cm},clip,width=\textwidth]{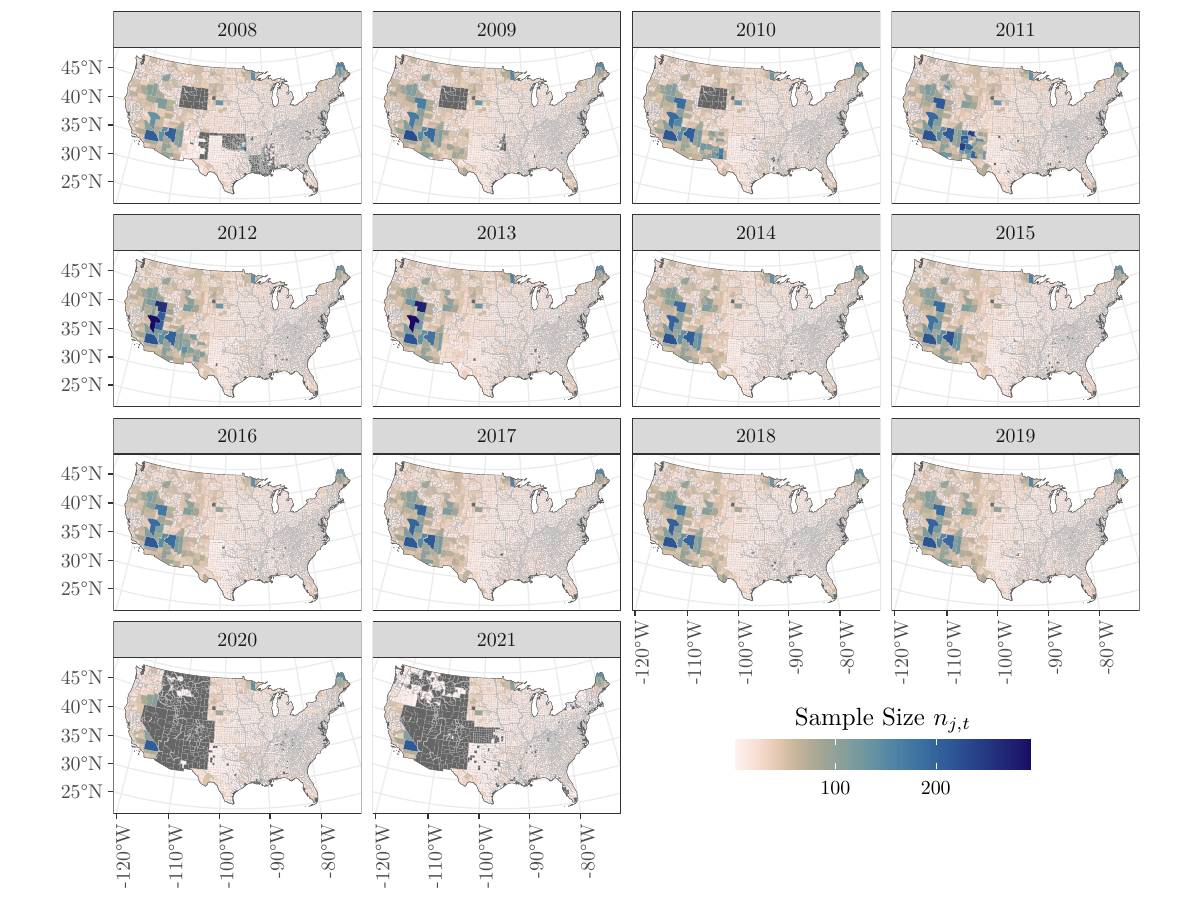}
    \caption{Number of FIA plots $n_{j,t}$ measured in county $j$ in year $t$. Gray counties indicate $n_{j,t} = 0$. Due to the FIA sampling design, larger counties are typically associated with greater sample sizes. FIA data used in this study were  from the FIA Database prior to 9/2024, since that time additional measurements for western states have been added for 2020 and 2021.}
    \label{fig:n_jt}
\end{figure}

The FIA program is responsible for monitoring the nation's forest resources through an NFI system of more than 300 thousand forest inventory plots spaced evenly across the CONUS \citep{bechtold_patterson_2005, WestfallFIA_2022}. Plot-level NFI data are collected by FIA on a rotating basis every 5 to 10 years, and include measurements of important forest parameters (e.g., tree diameter at breast height) which support estimation of attributes such as LFCD. Here, we use 593,368 FIA plot measurements collected across 3,108 counties in the CONUS from 2008 to 2021. Given the varying land areas of counties within the CONUS, and the panelized remeasurement intervals associated with the permanent FIA sample locations, between 0 and 294 inventory plots are measured within a given county in a given year as shown in Figure \ref{fig:n_jt}. Whereas these plot-level measurements would be used to calculate direct estimates in a traditional FH model, here, they serve as primary data to estimate average LFCD for each county and year.

Importantly, forest land is defined by the FIA program as land having at least 10\% tree canopy cover, or land that has had 10\% canopy cover in the past based on the presence of stumps or other evidence, and will be regenerated either artificially or naturally in the future. On non-forest inventory plots, FIA reports zero LFCD, even though there may as much as 10\% canopy cover on the plot. Therefore, if forest and non-forest plots are used for modeling, estimates of LFCD could be negatively biased. This is a known limitation and is discussed in \cite{Wiener2021} and \cite{Knott2023}. However, we use both forest and non-forest plot measurements in our subsequent analyses, and propose a model structure that can accommodate counties and years where few or no plot measurements are available, or where all plot measurements are identically zero, as is commonly the case in many arid, urban, or agricultural counties in the CONUS \citep{Shannon_2025}. 

\begin{figure}[ht!]
    \centering
    \includegraphics[trim={0.5cm 0cm 0.5cm 0.2cm},clip,width=\textwidth]{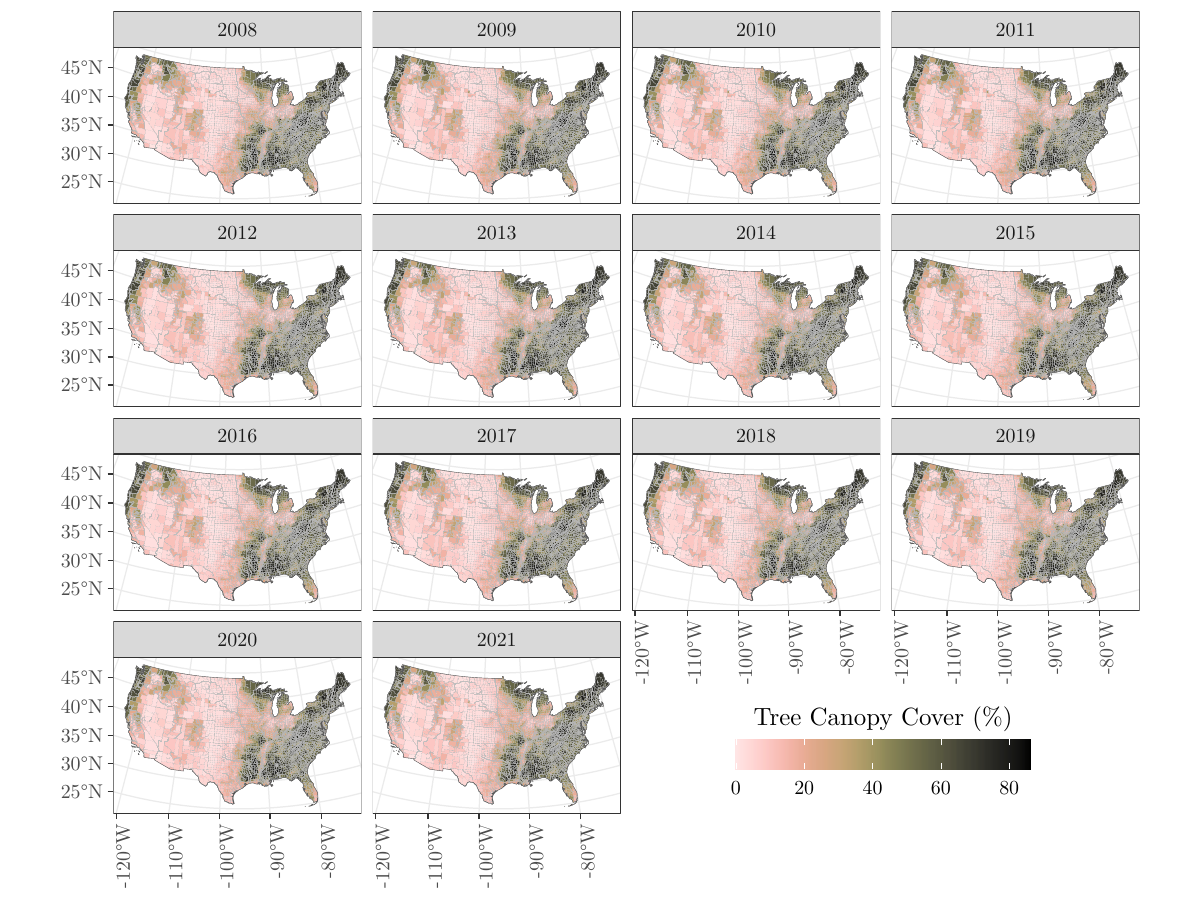}
    \caption{Percent tree canopy cover (TCC) averaged within each county and year.}
    \label{fig:tcc}
\end{figure}

To inform estimates of average LFCD for each county and year, we use a remotely sensed percent tree canopy cover (TCC) covariate that is produced by the USDA Forest Service as part of the National Land Cover Database (NLCD) \citep{tcc_methods_2023}. TCC is a remotely sensed data product derived from multispectral Landsat observations, and is available as annual raster images with a 30 m pixel resolution across the CONUS. Here, we calculate average \% TCC within each county and year to be used as a covariate in estimating average LFCD, which is shown in Figure \ref{fig:tcc}.

\subsection{Model}\label{sec:model}
We construct a Bayesian spatio-temporal SAE model applied to plot-level FIA LFCD data described in Section~\ref{sec:data}. Let $j = 1, \ldots, J$ index counties, $t = 1, \ldots, T$ index years and $i = 1, \ldots, n_{j,t}$ index FIA plot measurements, where $n_{j,t}$ is the number of FIA plot measurements available in county $j$ in year $t$. In some cases we may have $n_{j,t} = 0$. Additionally, let $y_{i,j,t}$ be the LFCD observation (Mg/ha) at FIA sample location $i$ in county $j$ measured in year $t$. Our goal, then, is to estimate the latent mean LFCD, denoted $\mu_{j,t}$, for county $j$ in year $t$. Let $P$ be the number of covariates that are collected into a vector of length $P + 1$ available for county $j$ in year $t$ as $\bx_{j,t}$, where the first element of $\bx_{j,t}$ represents an intercept term and is given as $1$. Additionally, we allow a subset of $Q \subseteq P$ many covariates to have a space-varying impact on the response, which are collected in a length $Q$ vector for county $j$ and year $t$ as $\tilde{\bx}_{j,t}$. The constraint that covariates in $\tilde{\bx}_{j,t}$ be a subset of those in $\bx_{j,t}$ is not required, but is common in models with spatially varying regression coefficients. For county $j$ in year $t$, the proposed model is then
\begin{equation}\label{eq:mod}
y_{i,j,t} = \underbrace{\bx_{j,t}^{\T} \bbeta_t + \tilde{\bx}_{j,t}^{\T} \bet_j + u_{j,t}}_{\mu_{j,t}} + \varepsilon_{i,j,t}, \quad i = 1, \ldots, n_{j,t},
\end{equation}
where $\varepsilon_{i,j,t}$ are independently distributed normal ($N$) residual error terms with time-specific variance, given as $N(0, \sigma^2_t)$. We then have that $\bbeta_t$ is a length $P + 1$ vector of temporally-varying regression coefficients corresponding to the intercept and covariates in $\bx_{j,t}$. Similarly, $\bet_{j}$ is a length $Q$ vector of space-varying regression coefficients corresponding to covariates in $\tilde{\bx}_{j,t}$. Finally, $u_{j,t}$ is a dynamically evolving spatio-temporal intercept term \citep{waller_1997} \citep[also see, e.g., Chapter~11 in][and references therein]{banerjee_carlin_gelfand}. We refer to (\ref{eq:mod}) as the ``full model'' in the remaining text. In Section~\ref{sec:model_comparison}, we consider a sub-model that excludes the space-varying regression term $\tilde{\bx}_{j,t} \bet_j$ from (\ref{eq:mod}), and perform model selection using the widely applicable information criterion (WAIC) \citep{gelman2013, Gelman2014, Vehtari2017}. The construction of terms $\bbeta_t$, $\bet_{j}$ and $u_{j,t}$ are described in the following sections. 

\subsubsection{Temporally-varying Regression Coefficients}\label{sec:beta}

The elements of $\bbeta_t$ in (\ref{eq:mod}) represent temporally-varying regression coefficients for covariates in $\bx_{j,t}$ \citep{West1997, stroud01, gel05}. Specifically, $\bbeta_t$ is modeled dynamically as
\begin{align}\label{eq:beta_t}
\bbeta_t &= \bbeta_{t - 1} + \bxi_t, \text{with} \\
\bxi_t &\indsim MVN \left(\bzero, \bSigma_{\xi} \right), \quad t = 1, \ldots, T,
\end{align}
where $\bxi_t$ are independently distributed evolution terms each following a mean zero multivariate normal ($MVN$) distribution with covariance matrix $\bSigma_{\xi}$, and $\bzero$ is a length $P+1$ vector of zeros. To complete the dynamic specification, we assign a multivariate normal prior to $\bbeta_0$ as $MVN \left(\bmu_0, \bSigma_0 \right)$, and an inverse-Wishart ($IW$) prior to $\bSigma_\xi$ as $IW \left(\nu_\xi, \bH_\xi \right)$. Modeling $\bbeta_t$ as in (\ref{eq:beta_t}) allows the effect of covariates in $\bx_{j,t}$ to have time-varying impact on the response, which varies dynamically over time according to the covariance structure in $\bSigma_\xi$. In other words, the distribution of $\bbeta_t$ has mean $\bbeta_{t - 1}$ and covariance matrix $\bSigma_\xi$, written as $MVN( \bbeta_{t-1}, \bSigma_\xi)$. 

\subsubsection{Space-varying Regression Coefficients}\label{sec:car}

The elements of $\bet_{j}$ in (\ref{eq:mod}) represent space-varying regression coefficients for covariates in $\tilde{\bx}_{j,t}$, with spatial dependency captured through neighborhood associations following a conditional autoregressive (CAR) model structure \citep{banerjee_carlin_gelfand}. Specifically, writing $\bet_{j} = (\eta_{1,j}, \ldots, \eta_{Q,j})^\top$ and collecting the $J$-many regression coefficients corresponding to covariate $q$ in a vector as $\bet^*_{q} = (\eta_{q,1}, \ldots, \eta_{q, J})^\top$, the CAR spatial structure for $\bet^*_{q}$ is specified as a multivariate normal distribution of the form
\begin{equation}\label{eq:bet_q}
\bet^*_{q} \sim MVN \left(\bzero, \tau^2_{\eta, q} \bQ(\rho_{\eta, q}) \right), \quad q = 1, \ldots, Q,
\end{equation}
where $\bzero$ is a length $J$ vector of zeros, $\tau^2_{\eta,q}$ is a scalar variance parameter, $\rho_{\eta, q}$ is a spatial correlation parameter ($0 < \rho_{\eta,q} < 1$), and $\bQ(\rho_{\eta, q})$ is a $J \times J$ correlation matrix of the form $\bQ(\rho_{\eta, q}) = (\bD - \rho_{\eta, q} \bW)^{-1}$. Here, $\bD$ is a $J \times J$ diagonal matrix whose $j^\text{th}$ diagonal element is the number of neighbors associated with county $j$, denoted as $d_j$. Counties are considered to be neighbors if they share a common boundary. Further, we require $d_j > 0$ for all $j$, i.e., no disconnected counties. Accordingly, $\bW$ is a fixed $J \times J$ binary spatial adjacency matrix with elements $\text{w}_{j,k} = 1$ if counties $j$ and $k$ are neighbors and $\text{w}_{j,k} = 0$ otherwise with $\text{w}_{j,j} = 0$ for all $j$. We specify $\tau^2_{\eta, q} \overset{ind}{\sim } IG(a_{\eta, q}, b_{\eta, q})$ and $\space\rho_{\eta, q} \overset{iid}{\sim} U(0,1)$ for $q = 1, \ldots, Q$ as priors, where $IG(a,b)$ denotes the inverse-Gamma distribution with shape $a$ and scale $b$, and $U(0,1)$ is the uniform distribution over the $(0,1)$ interval. Modeling $\bet^*_q$ as in (\ref{eq:bet_q}) introduces spatially-varying impact of covariate $q$ in $\tilde{\bx}_{j,t}$ on the response variable \citep{gelfandEtAl2003jasa}, with spatial variability determined by $\tau^2_{\eta, q}$ and strength of spatial correlation determined by $\rho_{\eta, q}$. 

\subsubsection{Dynamic Spatio-temporal Intercept}\label{sec:dynamic}

The spatio-temporally varying intercept term $u_{j,t}$ in (\ref{eq:mod}) is modeled as a dynamically evolving CAR spatial random effect. Specifically, we model $u_{j,t}$ dynamically as
\begin{equation}\label{u_j,t}
u_{j,t} = u_{j, t - 1} + \omega_{j,t},
\end{equation}
where $u_{j, 0} \equiv 0$ for all $j$. Then, collecting all $\omega_{j,t}$ for time $t$ as $\bomega_{t} = (\omega_{1,t}, \ldots, \omega_{J,t})^\top$, we specify a CAR spatial structure for $\bomega_{t}$ as
\begin{equation}\label{eq:omega_t}
\bomega_{t} \sim MVN \left(\bzero, \tau^2_{\omega, t} \bQ(\rho_{\omega}) \right),
\end{equation}
where the terms in (\ref{eq:omega_t}) are analogous to those described in Section~\ref{sec:car}. Here, we allow for time-specific spatial variance terms $\tau^2_{\omega,t}$. Similarly, time-specific spatial dependence parameters $\rho_{\omega,t}$ could be easily accommodated. Collecting all $J$-many $u_{j,t}$ at time $t$ in a vector as $\bu_t = (u_{1,t}, \ldots, u_{J,t})^\top$, their joint distribution is given as $\bu_t \sim MVN\left(\bu_{t - 1}, \tau^2_{\omega, t} \bQ(\rho_{\omega}) \right)$. We again assign inverse-gamma priors to variance parameters $\tau^2_{\omega,t}$ for $t = 1, \ldots, T$ as $IG(a_{\omega,t}, b_{\omega,t})$ and a uniform prior for $\rho_\omega$ as $U(0,1)$. 

\subsubsection{Priors and Likelihood}

To complete the Bayesian model specification, we assign inverse-gamma priors to variance parameters $\sigma^2_t, t = 1, \ldots, T$ as $IG(a_\sigma, b_\sigma)$. The joint posterior distribution for all parameters in model (\ref{eq:mod}) is then proportional to the product of the likelihood times priors, which is given as
\begin{align}\label{eq:posterior}
\begin{split}
&\prod^{T}_{t=1}\prod^J_{j=1} \prod^{n_{j,t}}_{i = 1} N\left(y_{i,j,t}\given \bx_{j,t}^\T \bbeta_t + \tilde{\bx}_{j,t}^\T \bet_{j} + u_{j,t},\, \sigma^2_{t}\right) \times \prod^{T}_{t=1} IG \left( \sigma^2_{t} \given a_\sigma, b_\sigma \right) \times\\
&\quad MVN \left( \bbeta_0 \given \bmu_0, \bSigma_0 \right) \times \prod_{t = 1}^T MVN\left(\bbeta_t \given \bbeta_{t-1}, \bSigma_\xi \right) \times IW \left( \bSigma_\xi \given \nu_\xi, \bH_\xi\right) \times\\
&\quad \prod^{Q}_{q=1} MVN\left(\bet^*_{q} \given \bzero, \tau^2_{\eta, q} \bQ(\rho_{\eta, q}) \right) \times \prod^{Q}_{q=1}  IG \left(\tau^2_{\eta, q} \given a_{\eta,q}, b_{\eta,q} \right) \times \prod^{Q}_{q=1} U\left(\rho_{\eta, q} \given 0, 1\right) \times\\
&\quad \prod^T_{t=1} MVN\left(\bu_{t} \given \bu_{t-1},\, \tau^2_{\omega, t} \bQ(\rho_{\omega})\right) \times \prod^{T}_{t=1} IG \left(\tau^2_{\omega, t} \given a_{\omega,t}, b_{\omega,t} \right) \times U\left(\rho_{\omega} \given 0, 1 \right).
\end{split}
\end{align}
Hyperparameters for prior distributions in (\ref{eq:posterior}) are chosen to offer vague specifications. Specifically, we set inverse-gamma shape parameters to 2, scale parameters to 100, $\nu_\xi = 10$, $\bH_\xi$ to be a $(P + 1) \times (P+1)$ diagonal matrix with diagonal elements equal to $100$, $\bmu_0 = \bzero$ and $\bSigma_0$ to be a $(P + 1) \times (P+1)$ diagonal matrix with diagonal elements equal to $100$. Finally, parameter inference is based on posterior samples from MCMC \citep[see, e.g.,][]{gelman2013} collected via Gibbs sampling updates for parameters whose full conditional distributions are readily available, and Metropolis algorithms for remaining parameters. Computing details, including Gibbs and Metropolis updates, are outlined in Section~\ref{sec:sampler_computing}. Results presented in Section~\ref{sec:results} are based on $M = 7,500$ posterior samples for model parameters, of which $2,500$ are retained and thinned as post burn-in samples. In subsequent analyses, we use TCC as the covariate, meaning $P = Q = 1$ and $\bx_{j,t}^{\T} = (1 : x_{TCC, j, t})$ while $\tilde{\bx}_{j,t}^{\T} = x_{TCC, j, t}$. 

\subsection{Direct Estimates}\label{sec:direct_estimates}

Traditionally, NFI-derived quantities of interest have been estimated using design-based direct estimates. Specifically, the direct estimate mean for $\mu_{j,t}$ is calculated as
\begin{equation}\label{eq:direct_mean}
\hat{\mu}_{j,t} = \frac{1}{n_{j,t}} \sum_{i = 1}^{n_jt} y_{i,j,t}
\end{equation}
with associated estimate variance
\begin{equation}\label{eq:direct_se}
\hat{\sigma}^2_{j,t} = \frac{1}{n_{j,t}(n_{j,t} - 1)} \sum_{i = 1}^{n_jt} (y_{i,j,t} - \hat{\mu}_{j,t})^2.
\end{equation}
These direct estimates serve as the baseline to which we will compare the model estimates in Section \ref{sec:results}. Maps of the direct estimate means (\ref{eq:direct_mean}) and estimate variances (\ref{eq:direct_se}) are shown in Figures~\ref{fig:direct_mean} and \ref{fig:direct_se}, respectively. 

Design-based estimates are customarily motivated from sampling of finite populations. Population parameters are accessible without error if all population units are observed. Randomness is imposed through a randomized sampling design used to select population units into a sample \citep{kish1965book, Sarndal1978, Gregoire1989}. However, as noted in Section~\ref{sec:introduction}, these direct estimates are often unavailable. Specifically, $\hat{\mu}_{j,t}$ is unavailable when $n_{j,t} = 0$, and $\hat{\sigma}^2_{j,t}$ is unavailable when $n_{j,t} = 0$, $n_{j,t} = 1$, or when $y_{i,j,t} = y_{k,j,t}$ for all $1 \leq i \neq k \leq n_{j,t}$. For this last case, in the NFI data, sparsely forested regions typically have plot measurements equaling $0$, which precludes reporting $\hat{\sigma}^2_{j,t}$. Design-based inference is unfeasible here so we adopt Bayesian modeling to ensure full propagation of uncertainty in imputation of missing data \citep{little2004jasa, Ghosh:2012md, banerjee2023finite}. 

\subsection{Trends}\label{sec:trends}

To better understand how LFCD has changed over time, estimates of $\mu_{j,t}$ may be used to derive estimated trends measured in Mg/ha/year for each county $j = 1, \ldots, J$. Specifically, given $M$ posterior samples of $\mu_{j,t}$, indexed by $\mu_{j,t}^m$ for $m = 1, \ldots M$, we derive an estimated trend value for county $j$, denoted as $\theta_j$, whose posterior inference is available via
\begin{equation}\label{eq:trend}
\theta^{m}_j = \frac{\sum^T_{t=1}\left(t - \bar{t}\right)\left(\mu^{m}_{j,t} - \bar{\mu}^{m}_{j}\right)}{\sum^T_{t = 1}\left(t - \bar{t}\right)^2}, \quad m = 1, \ldots M,
\end{equation}
where $\bar{t} = \frac{1}{T} \sum_{t = 1}^T t$ and $\bar{\mu}^m_{j} = \frac{1}{T} \sum_{t = 1}^T \mu^m_{j,t}$. Counties with significant trend values are then determined by observing if the 95\% credible interval for $\theta_j$ overlaps zero. 

\subsection{Simulation}\label{sec:sim_methods}

To assess the performance of the proposed model (\ref{eq:mod}) and compare model fit metrics including bias, root mean square error (RMSE), coverage rate, and coverage interval widths against a direct estimator (\ref{eq:direct_mean}), a simulated population was constructed over a dense grid across the CONUS as in \cite{Shannon_2025} to mimic the spatial distribution and characteristics of LFCD as observed in the real FIA and TCC data. From the simulated population, $R = 100$ replicates were sampled at the same intensity as the observed FIA data as shown in Figure \ref{fig:n_jt}. Estimates from the full model (\ref{eq:mod}) were then obtained for each of the $R = 100$ replicates, along with direct estimates, and measures of their average bias, RMSE, coverage rate, and coverage interval widths over the replicates were then compared. Methods for computing bias, RMSE, coverage, and coverage interval widths are given in Section~\ref{sec:bias_rmse_coverage}.

\section{Results}\label{sec:results}

\subsection{Model Comparison}\label{sec:model_comparison}

The full model (\ref{eq:mod}) and sub-model (excluding the space-varying regression term $\tilde{\bx}_{j,t} \bet_j$) described in Section~\ref{sec:model} were both fit to the FIA plot-level data described in Section~\ref{sec:data}, and were compared using WAIC as implemented in the \texttt{loo} \texttt{R} package \citep{Watanabe2010,loo2024}. Specifically, $\widehat{\text{elpd}}_{\text{WAIC}}$ is the expected log pointwise predictive density, $\widehat{p}_{\text{WAIC}}$ is the effective number of parameters, and WAIC is the calculated as $-2 \times \widehat{\text{elpd}}_{\text{WAIC}}$. Here, lower WAIC values indicate better model performance. For more details on model selection via WAIC, see \cite{Vehtari2017}. Based on WAIC estimates provided in Table~\ref{tab:waic}, we select the full model as the preferred model, and results presented in the following sections represent full model estimates. 

\begin{table}[ht!]
\caption{Estimates for WAIC and associated statistics comparing the full and sub-model. The last row holds the estimated $\widehat{\text{elpd}}_{\text{WAIC}}$ difference with the ``best'' fitting (full) model and associated standard error $\widehat{\tau}_{\text{diff}}$ in parentheses.} 
\begin{center}
\begin{tabular}{lccc}
\toprule
&\multicolumn{2}{c}{Candidate models} \\
\cmidrule(lr){2-3} 
 & Full Model & Sub-Model\\
\midrule
$\widehat{\text{elpd}}_{\text{WAIC}}$ &-2866326.5 (1811.4)&-2866380.6 (1813.7)\\
$\widehat{p}_{\text{WAIC}}$ &4814.3 (44.1)&4459.2 (43.9)\\
WAIC &5732652.9 (3622.7)&5732761.2 (3627.3)\\
$\widehat{\text{elpd}}_{\text{diff}}$ & 0 (0)& -54.1 (40.2)\\
\bottomrule
\end{tabular}
\label{tab:waic}
\end{center}
\end{table} 

\subsection{FIA Data Analysis}

\begin{figure}[ht!]
    \centering
    \includegraphics[trim={0.5cm 0cm 0.5cm 0.2cm},clip,width=\textwidth]{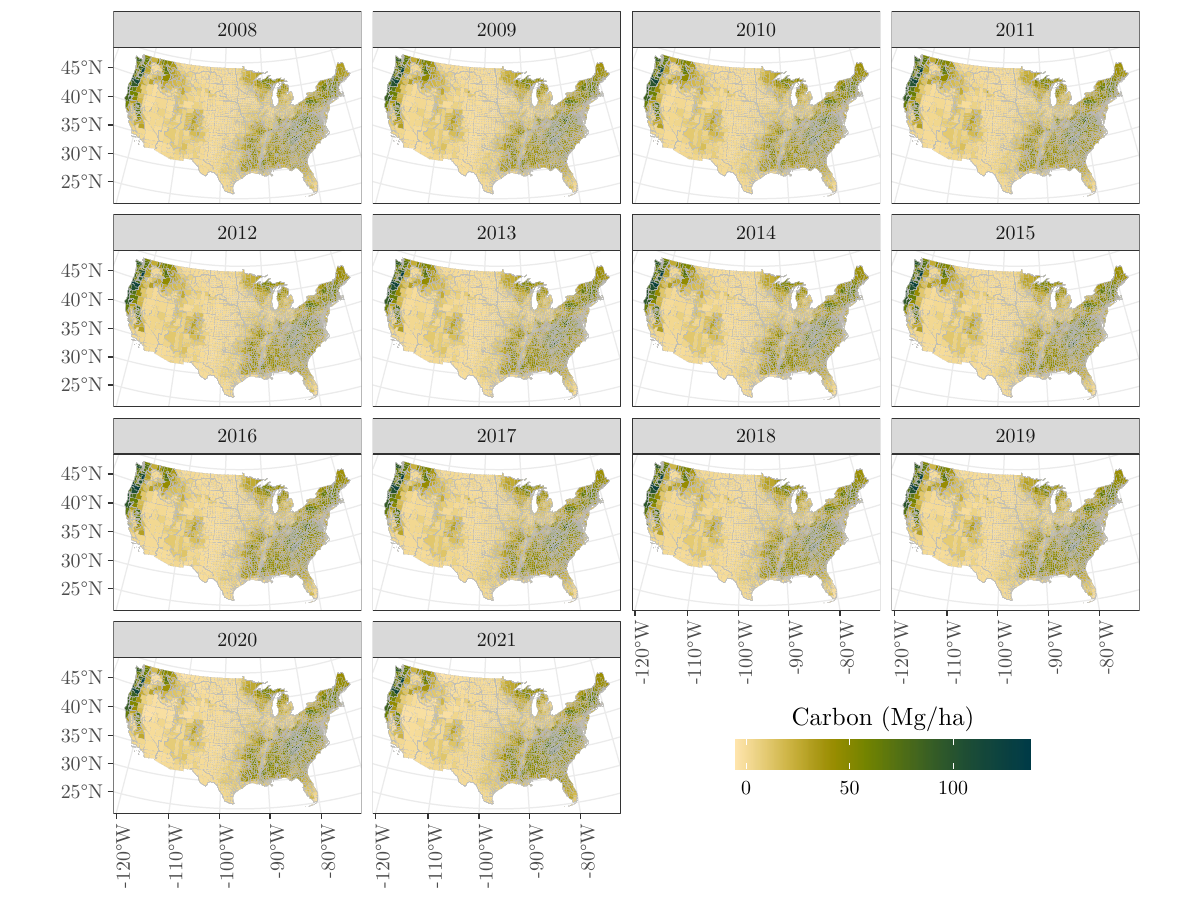}
    \caption{Posterior mean values of live forest carbon density ($\mu_{j,t}$) from the full model.}
    \label{fig:mu_mean}
\end{figure}

Fitting the full model (\ref{eq:mod}) to the FIA plot-level data is described in Section~\ref{sec:data}. Posterior distributions of the average LFCD $\mu_{j,t}$ are obtained and posterior means are shown in Figure~\ref{fig:mu_mean}, while posterior standard deviations are shown in Figure~\ref{fig:mu_sd}. Posterior summaries for temporally-varying regression coefficients $\bbeta_t$ (Figure~\ref{fig:beta}), space-varying regression coefficients $\bet_j$ (Figure~\ref{fig:svc}), dynamic spatio-temporal intercept terms $u_{j,t}$ (Figure~\ref{fig:u}), and time specific variance terms $\tau^2_{\omega,t}$ (Figure~\ref{fig:tau_sq_w}) and $\sigma^2_{t}$ (Figure~\ref{fig:sigma_sq}) are given in Section~\ref{sec:supporting_figs}. Posterior mean and 95\% credible interval bounds for $\rho_{\eta, TCC}$ were 0.993 (0.983, 0.998), and for $\rho_{\omega}$ they were 0.998 (0.997, 0.999). 

\begin{figure}[ht!]
    \centering
    \includegraphics[width=\textwidth]{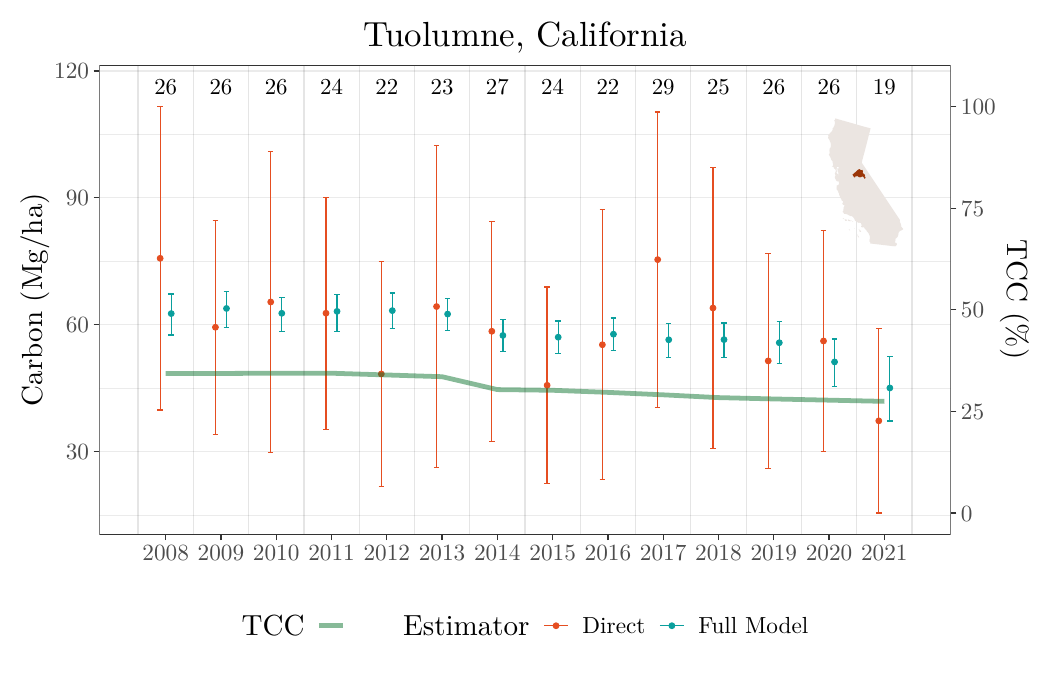}
    \caption{Posterior mean and 95\% credible intervals of live forest carbon density ($\mu_{j,t}$) from the full model for Tuolumne County, California, compared to direct estimate means ($\hat{\mu}_{j,t}$) and 95\% confidence intervals over time. Percent tree canopy cover (TCC) is overlayed and is measured on the right axis. Observed sample sizes ($n_{j,t}$) are displayed across the top row. An inset map shows the location of Tuolumne County, California, within the state. The abrupt drop in TCC and full model estimates in 2013 coincides with the catastrophic Rim fire.}
    \label{fig:tuolumne}
\end{figure}

\begin{figure}[ht!]
    \centering
    \includegraphics[width=\textwidth]{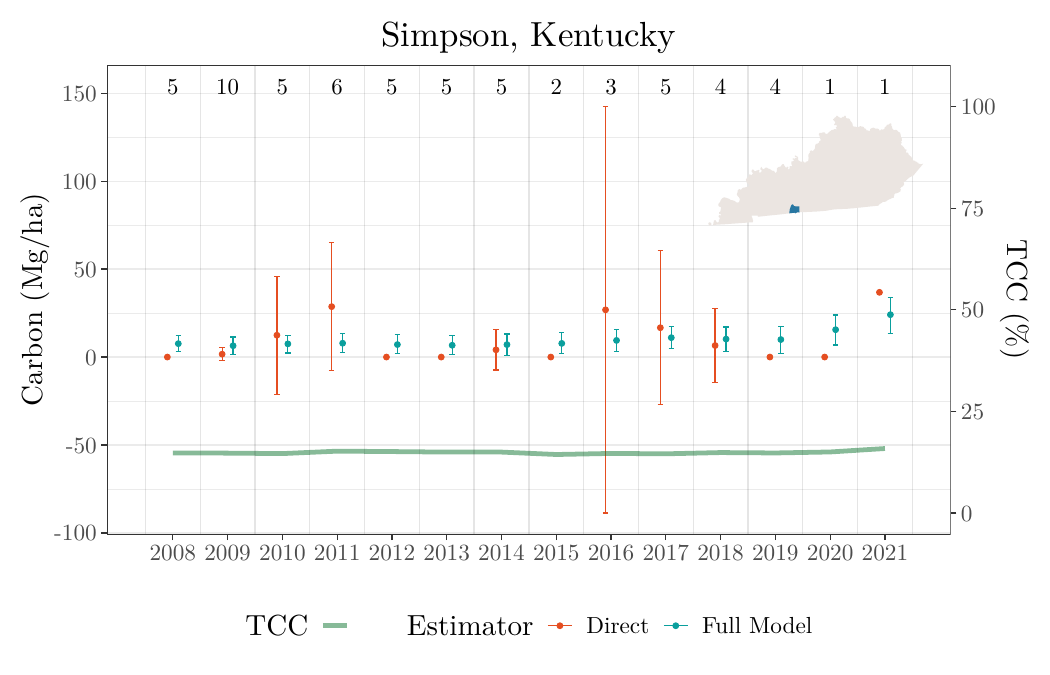}
    \caption{Posterior mean and 95\% credible intervals of live forest carbon density ($\mu_{j,t}$) from the full model for Simpson County, Kentucky, compared to direct estimate means ($\hat{\mu}_{j,t}$) and 95\% confidence intervals. TCC is overlayed and is measured on the right axis. Observed sample sizes ($n_{j,t}$) are displayed across the top row. An inset map shows the location of Simpson County, Kentucky, within the state. Direct estimate variances are unavailable in years when $n_{j,t} = 1$ or when all plot-level measurements are identical.}
    \label{fig:simpson}
\end{figure}

Model estimates of mean LFCD for individual counties are readily available, and visualizing these estimates over time provides important insights into forest carbon dynamics for small areas. Next, we highlight some qualities of the model and present the full model posterior estimates of $\mu_{j,t}$ for two selected counties, which we compare with direct estimates and show in Figures~\ref{fig:tuolumne} and \ref{fig:simpson}. In Figure~\ref{fig:tuolumne}, model estimates for Tuolumne County, California are shown, where the variability in model estimates is greatly reduced compared to the direct estimates, and changes in mean LFCD are more heavily influenced by TCC and exhibit more gradual changes over time. Still, model estimates more clearly capture abrupt changes in LFCD as indicated by changes in TCC, while the direct estimates remain variable and uncertain. Moreover, the abrupt drop in TCC in 2013 coincides with the catastrophic Rim fire, which burned 257,314 acres, including 154,530 acres of forest lands and is better captured by the full model estimates. 

In Figure~\ref{fig:simpson}, full model and direct estimates of $\mu_{j,t}$ for Simpson County, Kentucky are compared over time, and highlight the ability of the proposed model to leverage data that would otherwise lead to missing direct estimates in traditional design-based frameworks. Specifically, direct estimates for the error variance (\ref{eq:direct_se}) associated with the direct estimate mean are missing in years when $n_{j,t } = 1$ (2020 and 2021) as well as in years when all plot-level measurements $y_{i,j,t}$ are identically zero (2008, 2012, 2013, 2015, and 2019). In the current model setting, these data are used to inform estimates of $\mu_{j,t}$, and allow for model estimates to more clearly reflect information offered by the available data. In the NFI data setting, direct estimates of mean LFCD for small areas and time scales are typically of little value. The proposed model optimally pools information from spatially and temporally proximate direct estimates and TCC to provide a more informed estimate. 

\begin{figure}[ht!]
    \centering
    \includegraphics[width=\textwidth]{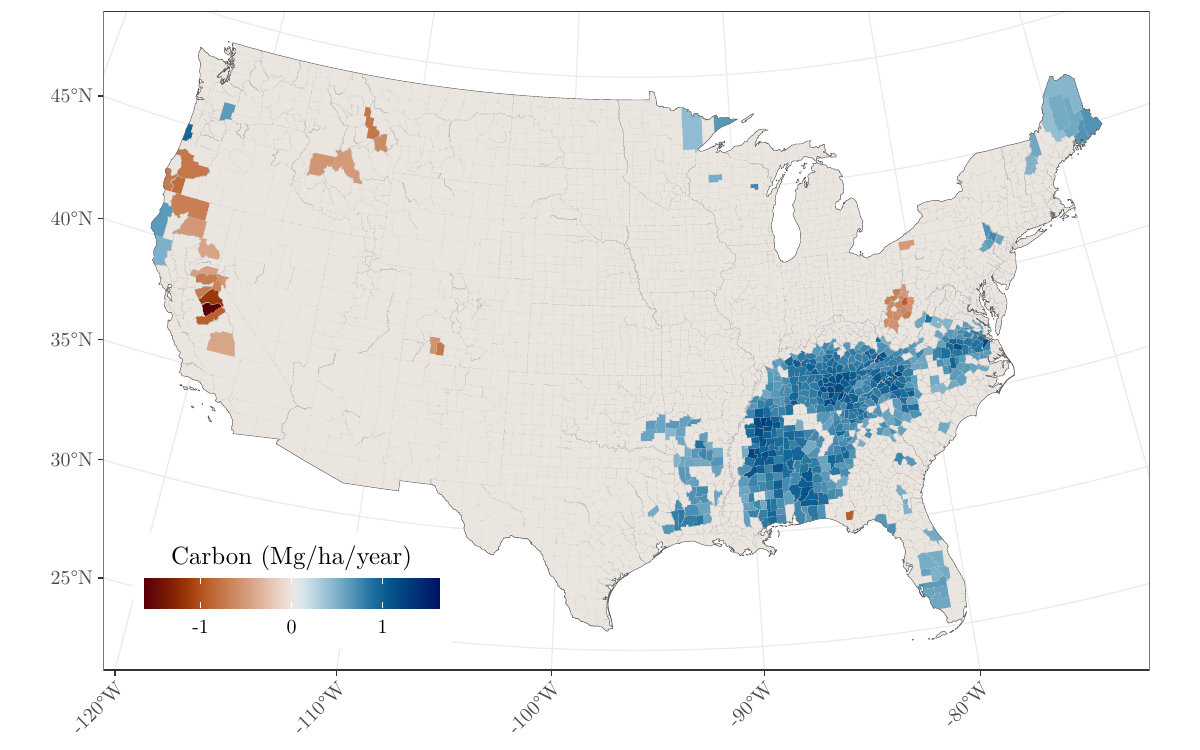}
    \caption{Significant live forest carbon density trends ($\theta_j$) measured in Mg/ha/year. ``Significant'' indicates that the 95\% credible interval for $\theta_j$ does not overlap zero. Non-significant trends shown in Figure~\ref{fig:trend} have been set equal to zero.} 
    \label{fig:sig_trend}
\end{figure}

Posterior means for carbon density trends $\theta_j$ are shown in Figure~\ref{fig:trend}, with significant trends shown in Figure~\ref{fig:sig_trend}. Figure~\ref{fig:sig_trend} highlights varied patterns of carbon dynamics across the CONUS, whereby LFCD is seen to be significantly increasing in areas including Northern Maine, the Southeast, and coastal western areas, while the Sierra Nevadas, Northern Rockies, and parts of the Appalachias exhibit decreasing trends. These results are comparable to those reported in \cite{Shannon_2025}, and reflect broader patterns relating to carbon dynamics driven by disturbance (e.g., fire, harvest, storm, insect/disease) and change in structure over time.

\subsection{Simulation Study}\label{sec:simulationOverview}

\begin{figure}[ht!]
    \centering
    \includegraphics[width=\textwidth]{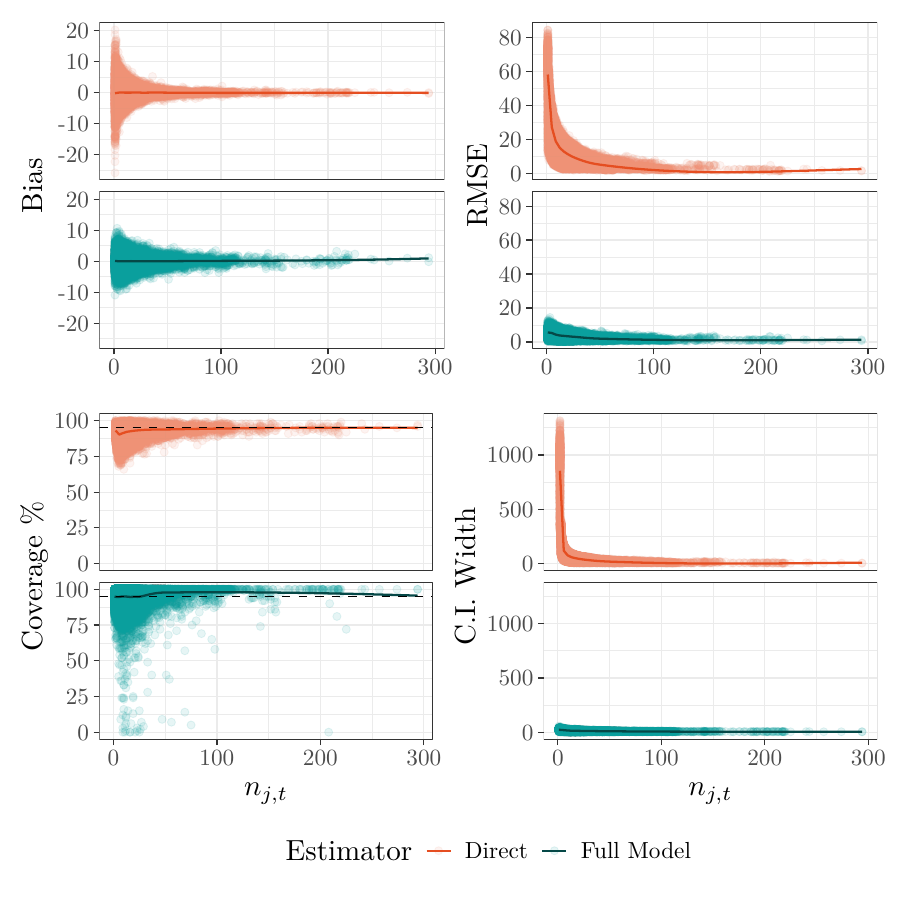}
    \caption{Average measures of bias, root mean square error (RMSE), coverage percentage, and coverage interval widths for the full model and direct estimator arranged according to sample size $n_{j,t}$. Each point represents the mean metric value for estimating $\mu_{j,t}$ averaged over $R = 100$ simulated population replicates. A loess line is added to each panel to indicate general patterns, and a 95\% threshold for the coverage percentages is represented with a dashed line.}
    \label{fig:sim_results}
\end{figure}

As detailed in Section~\ref{sec:sim_methods}, full model (\ref{eq:mod}) estimates, along with direct estimates, were obtained from $R = 100$ sample replicates from the simulated population. Measures of bias, RMSE, coverage rate, and precision for model versus direct estimates of $\mu_{j,t}$ were averaged over the replicates, with details of their calculation provided in Section~\ref{sec:bias_rmse_coverage}. Figure~\ref{fig:sim_results} compares the full model (\ref{eq:mod}) and direct estimate accuracy and precision for all $J \times T$ many $\mu_{j,t}$ estimates, which are ordered according to their corresponding sample size $n_{j,t}$, as shown in Figure~\ref{fig:n_jt}. Here, it is clear that both the full model and direct estimates may be biased for small sample sizes, with RMSE increasing for the smallest sample sizes. However, the overall bias and RMSE are greater for the direct estimate compared to the full model. Additionally, both the direct and full model estimates achieve similar coverage percentages, while the coverage interval widths are much larger for the direct estimates compared to the full model for small samples sizes. These results highlight the overall trend that both accuracy and precision are decreased when data are limited, with direct estimates exhibiting poorer accuracy and precision overall. 

\section{Discussion}\label{sec:Discussion}
Here, we present a Bayesian spatio-temporal SAE model applied to NFI measurements of LFCD across the CONUS. Compared to direct estimates, as well as SAE methods such as the FH model which rely on direct estimates, the proposed model more efficiently leverages available NFI data to inform estimates of status and trends for small areas. Specifically, in place of direct estimates, plot-level measurements are directly modeled using available covariates, and spatial, temporal, and spatio-temporal correlations in random effects and among responses are incorporated to improve estimation. When used to analyze FIA measurements of LFCD, the model delivers estimates that reflect changes in TCC and correlations with spatially and temporally proximate data, with reduced uncertainty compared to direct estimates (Figures~\ref{fig:tuolumne} and \ref{fig:simpson}). Furthermore, the proposed model informs the estimates using data that would otherwise be unusable in design-based estimators (Figure~\ref{fig:simpson}) and hence FH models. These estimates are easily extended to derive estimates of trends in LFCD over time, and highlight important forest carbon dynamics across the CONUS (Figure~\ref{fig:sig_trend}). When compared in a simulation setting, the proposed model exhibits improved accuracy and precision over the direct estimator, especially for instances when samples sizes are small (Figure~\ref{fig:sim_results}). 

A major advantage of the proposed model is the ability to directly use all available plot-level NFI data. As noted in Section~\ref{sec:direct_estimates}, when the sample size $n_{j,t}$ equals $1$, the associated error variance for the direct estimate mean (\ref{eq:direct_se}) is unavailable, meaning these data are essentially ignored in traditional SAE FH models. Additionally, when $n_{j,t} > 1$ and all plot measurements are identically zero, (\ref{eq:direct_se}) will also be unavailable, meaning potentially informative data are again omitted in these cases (see Figure~\ref{fig:simpson}). In the FIA data setting described in Section~\ref{sec:data}, out of 593,368 plot measurements used to fit the proposed model, 98,574 (16.6\%) result in missing associated direct estimates, and would therefore be omitted when using methods such as the FH model. Similarly, of $J \times T =$ 43,512 many $\mu_{j,t}$ county-year combinations for which we seek inference, 8,764 (20.1\%) have missing associated direct estimates due to $n_{j,t}$ being equal to 1, or due to all plot measurements being identically zero. Hence, the proposed model more efficiently uses the available data, and in the case of SAE for sparse and costly NFI data, these efficiencies are especially advantageous. 

Although the proposed model demonstrates a certain improvement over design-based direct estimation, a number of limitations persist. First, in cases where the observed sample size is small, biases in model estimates are more pronounced (Figure~\ref{fig:sim_results}). However, these impacts might be mitigated if proximate counties and times contain more observations to inform model estimates. Second, given the large and dynamic nature of the system we are modeling, there are many parameters to estimate. As outlined in Section~\ref{sec:sampler_computing}, closed-form expressions for posterior distributions are available when the response variable is Gaussian and therefore can be updated using computationally efficient Gibbs steps. The proposed efficient sampler can also be applied to binomial response variables when using a P{\'o}lya-gamma augmentation \citep[see][]{Polson_2013}. However, such efficient parameter updating methods might not be available for distributions commonly chosen for count or composition response variables, which are commonly encountered in NFI data. Future work could focus on developing efficient models for these type of data and for multivariate responses.

\section{Conclusion}\label{sec:Conclusion}

The use of SAE methods to estimate status and trends of forest parameters from NFI data represents an important opportunity to more fully use valuable NFI data efficiently and provide reliable and high-quality estimates with appropriate measures of uncertainty. Practitioners, resource managers, and policy-makers should benefit from these informed estimates and methods, which can be applied to a variety of data and model settings. Continued efforts to improve SAE models for NFI data should remain an important focus moving forward, as economic, environmental, and societal drivers continue to alter forests across varied scales. 

Given the high cost and effort to conduct NFIs, the ability to efficiently use all available data to estimate status and trends in forest parameters for small areas is critical for monitoring of forest resources and sustainable management of these resources. SAE models, such as the model proposed here, provide clear paths toward reliable and high-quality estimates to improve efforts to monitor and manage forests.

\section*{Funding}\label{sec:Funding}

This work was supported by USDA Forest Service (FS), USDA FS/NCASI Partnership on Small Area Estimation, National Science Foundation (NSF) DEB-2213565 and DEB-1946007, and National Aeronautics and Space Administration (NASA) CMS Hayes-2023. The findings and conclusions in this publication are those of the authors and should not be construed to represent any official USDA or US Government determination or policy. 

\clearpage

\makeatletter
\renewcommand \thesection{S\@arabic\c@section}
\renewcommand\thetable{S\@arabic\c@table}
\renewcommand \thefigure{S\@arabic\c@figure}
\renewcommand \theequation{S\@arabic\c@equation}
\makeatother

\section*{Supporting information}\label{sec:supporting}

\section{Supplementary Figures}\label{sec:supporting_figs}

\begin{figure}[ht!]
    \centering
    \includegraphics[trim={0.5cm 0cm 0.5cm 0.2cm},clip,width=\textwidth]{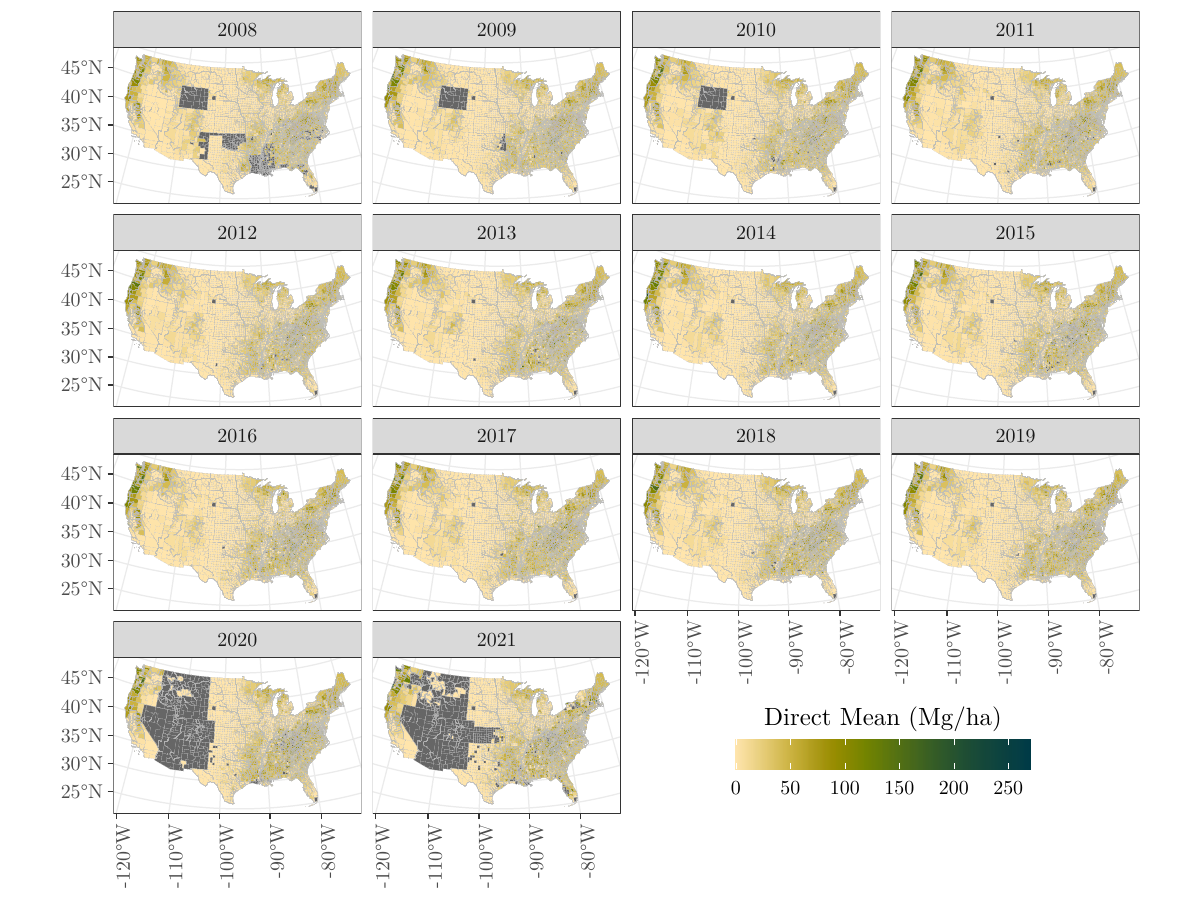}
    \caption{Direct estimates of mean LFCD ($\hat{\mu}_{j,t}$) (Mg/ha). Gray counties indicate that the direct estimate is missing due to $n_{j,t} = 0$. }
    \label{fig:direct_mean}
\end{figure}

\begin{figure}[ht!]
    \centering
    \includegraphics[trim={0.5cm 0cm 0.5cm 0.2cm},clip,width=\textwidth]{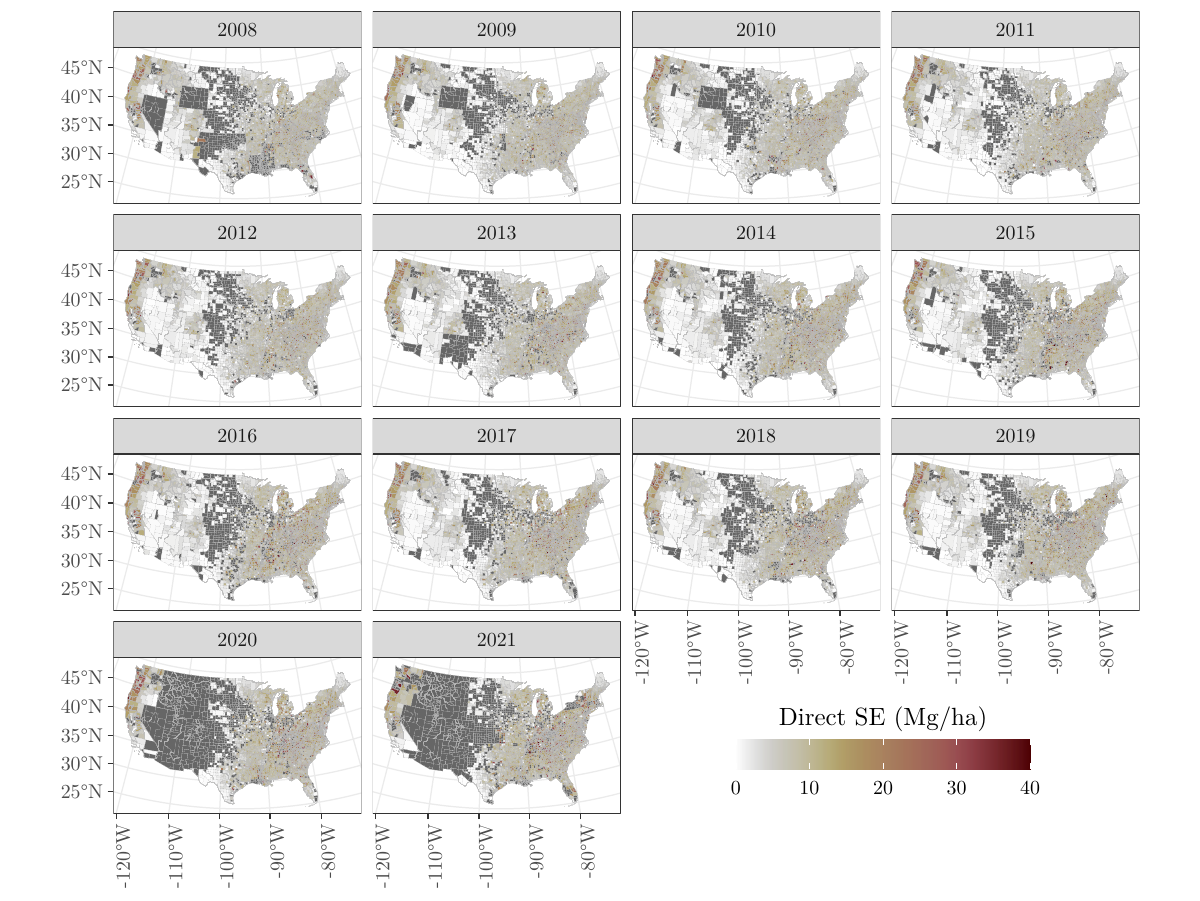}
    \caption{Direct estimate standard error ($\hat{\sigma}^2_{j,t}$) corresponding to the direct estimate mean in Figure~\ref{fig:direct_mean}. Gray counties indicate the direct estimate standard error is missing due to $n_{j,t} = 0$, $n_{j,t} = 1$, or due to all plot-level measurements being identical in the county and year.}
    \label{fig:direct_se}
\end{figure}

\begin{figure}[ht!]
    \centering
    \includegraphics[trim={0.5cm 0cm 0.5cm 0.2cm},clip,width=\textwidth]{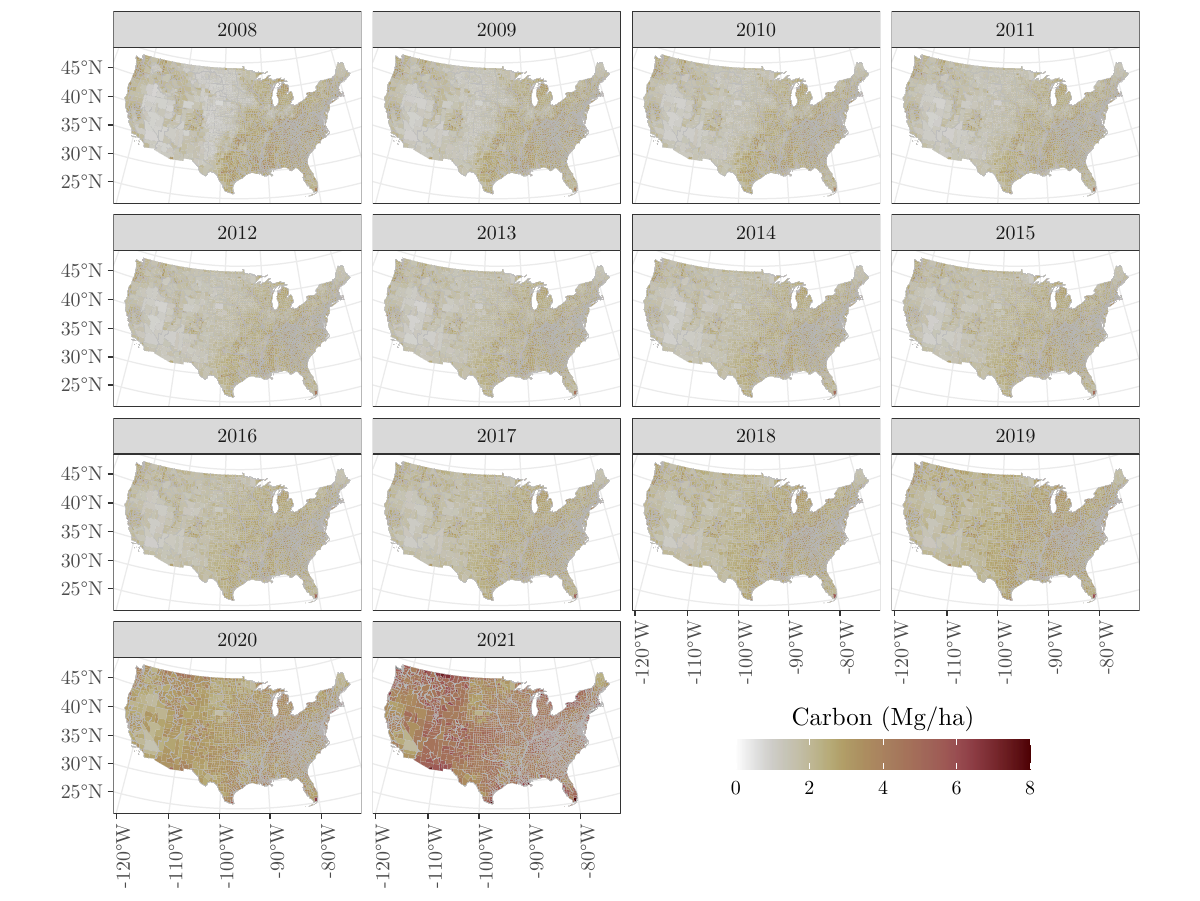}
    \caption{Posterior standard deviation of mean LFCD ($\mu_{j,t}$) from the full model (Mg/ha). Increasing values over time are influenced by decreased sample sizes in later years, as shown in Figure~\ref{fig:n_jt}.}
    \label{fig:mu_sd}
\end{figure}

\begin{figure}[ht!]
    \centering
    \includegraphics[width=\textwidth]{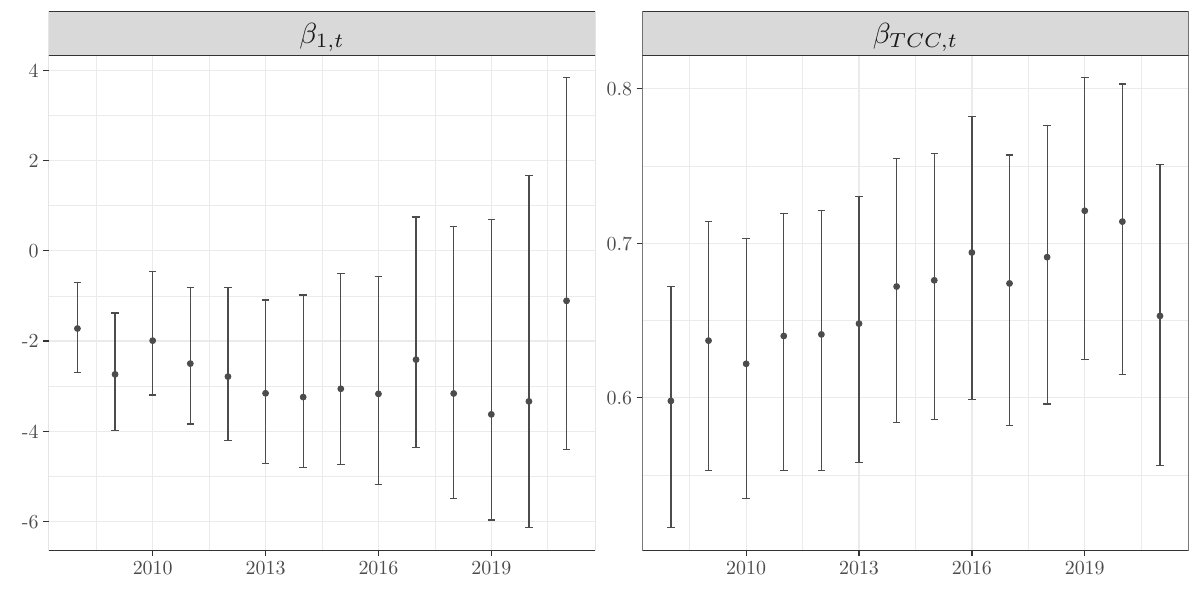}
    \caption{Full model posterior mean and upper and lower 95\% credible interval values for the temporally-varying regression coefficient $\bbeta_t = (\beta_{1,t}, \beta_{TCC, t})^\T$.}
    \label{fig:beta}
\end{figure}

\begin{figure}[ht!]
    \centering
    \includegraphics[width=\textwidth]{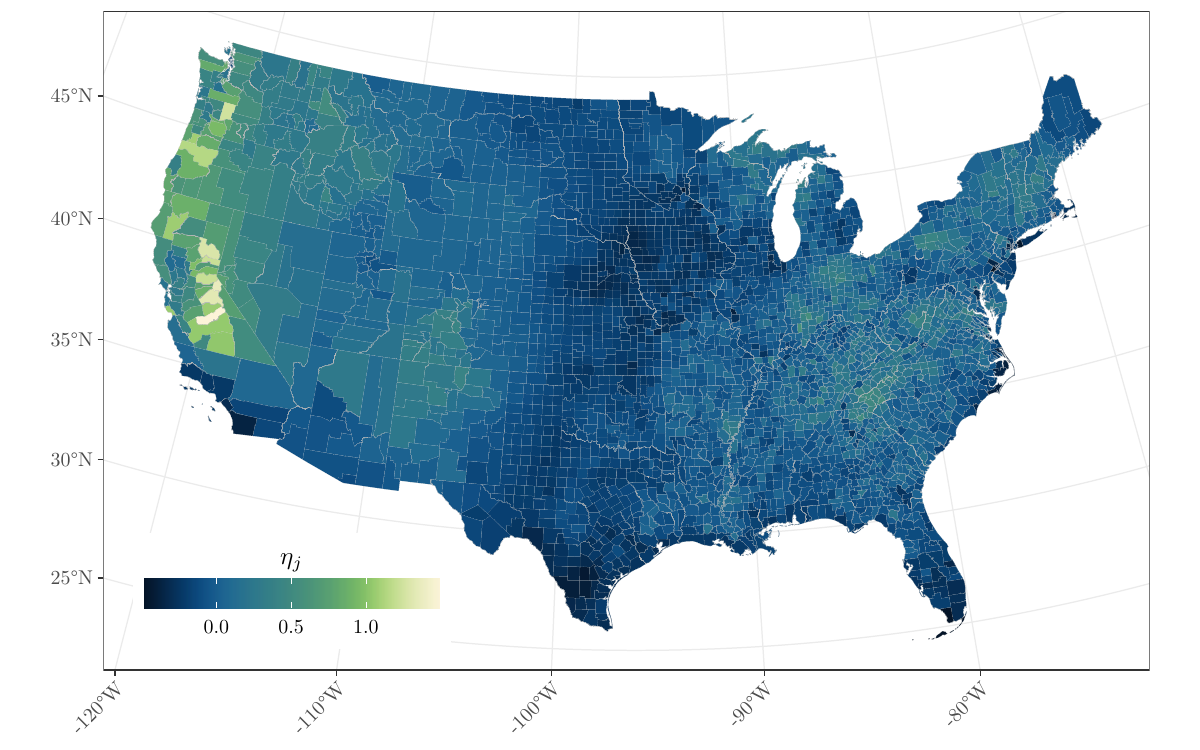}
    \caption{Full model posterior mean values for the space-varying regression coefficient $\eta_j$. The impact of TCC on the response displays clear spatial variability over the CONUS, where increased TCC in counties in the far west and parts of the east are associated with greater increases in mean LFCD.}
    \label{fig:svc}
\end{figure}

\begin{figure}[ht!]
    \centering
    \includegraphics[trim={0.5cm 0cm 0.5cm 0.2cm},clip,width=\textwidth]{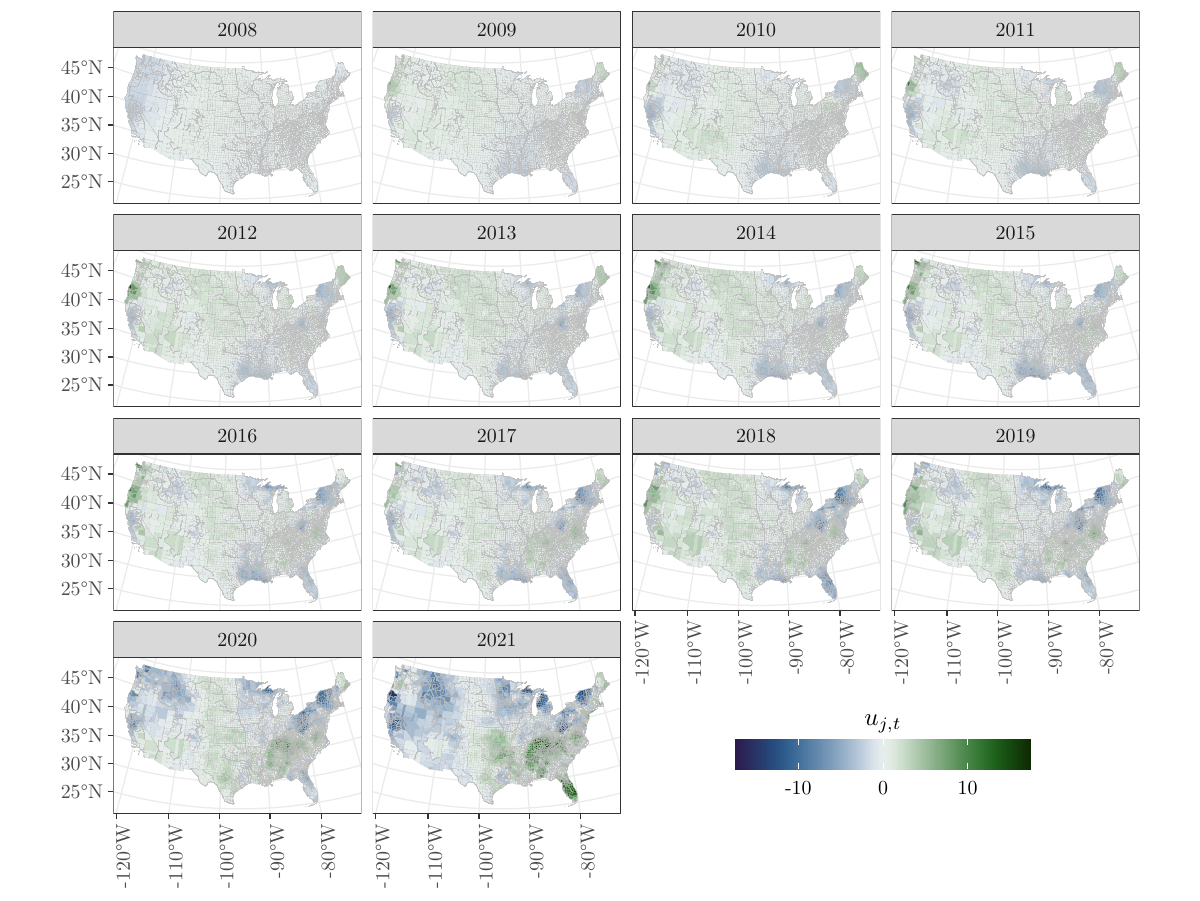}
    \caption{Full model posterior mean values of the dynamically evolving space-varying intercept term $u_{j,t}$.}
    \label{fig:u}
\end{figure}

\begin{figure}[ht!]
    \centering
    \includegraphics[width=\textwidth]{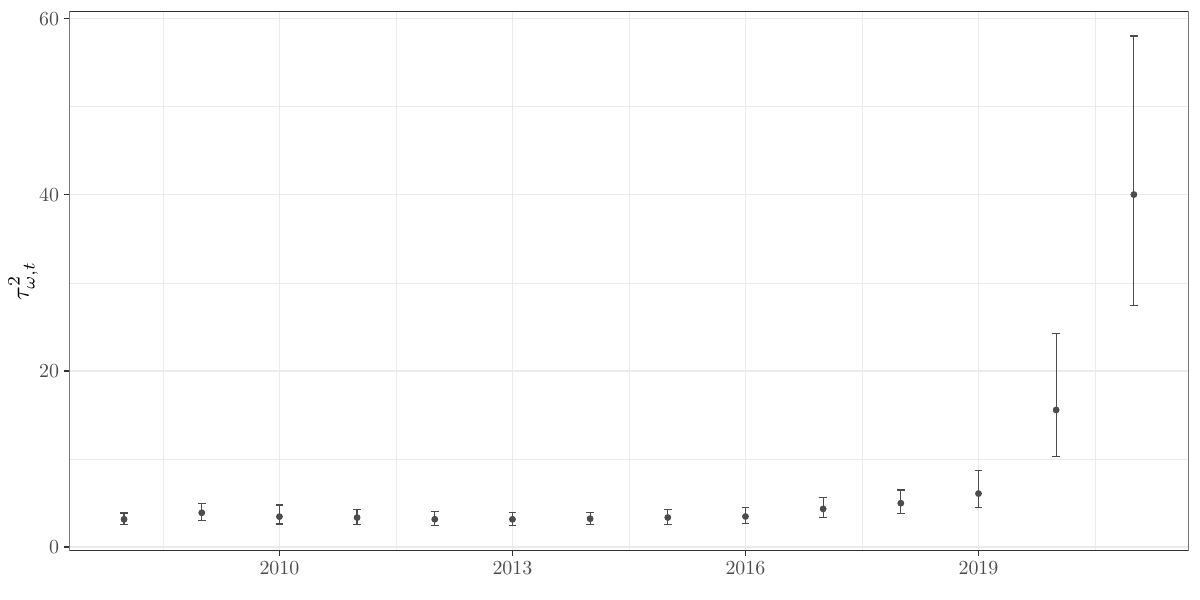}
    \caption{Full model posterior mean and upper and lower 95\% credible interval values for the time specific spatial variance term $\tau^2_{\omega,t}$.}
    \label{fig:tau_sq_w}
\end{figure}

\begin{figure}[ht!]
    \centering
    \includegraphics[width=\textwidth]{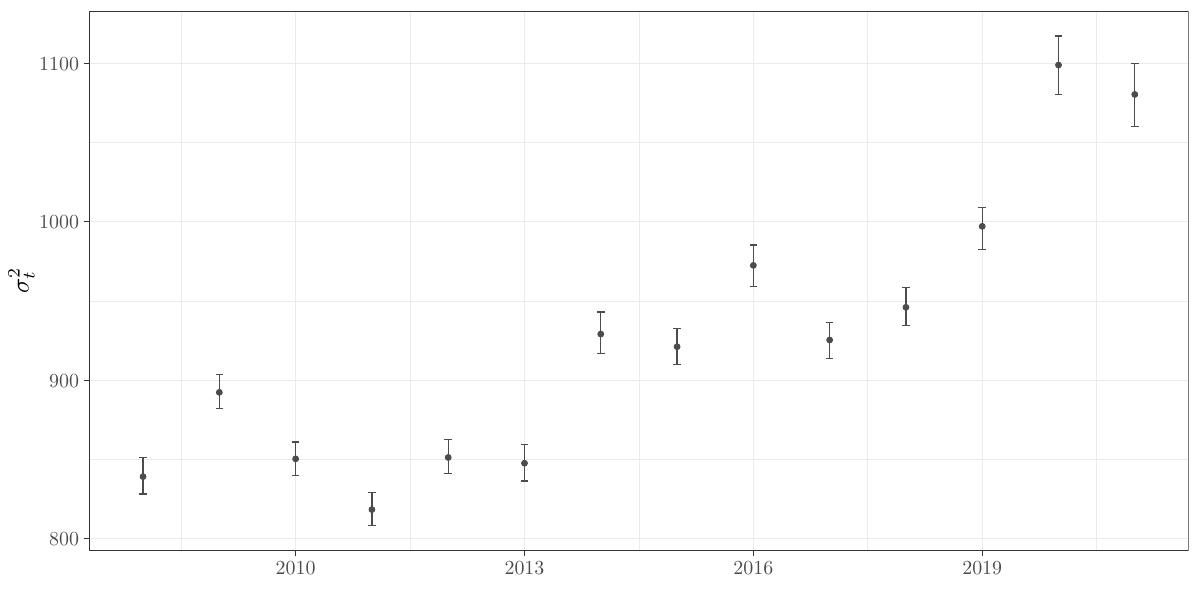}
    \caption{Full model posterior mean and upper and lower 95\% credible interval values for the time specific residual variance term $\sigma^2_{t}$.}
    \label{fig:sigma_sq}
\end{figure}

\begin{figure}[ht!]
    \centering
    \includegraphics[width=\textwidth]{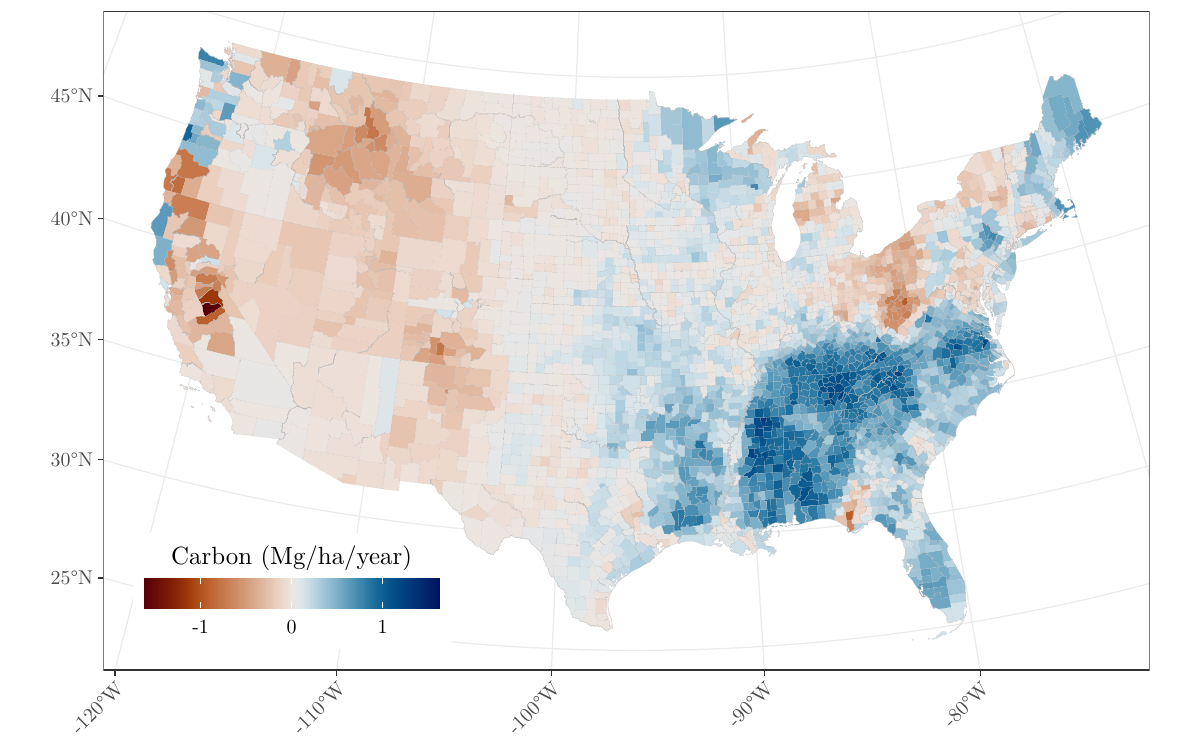}
    \caption{Estimated mean LFCD trends ($\theta_j$) measured in Mg/ha/year.}
    \label{fig:trend}
\end{figure}

\clearpage

\section{Bias, RMSE, Coverage and Precision}\label{sec:bias_rmse_coverage}

From the simulated population constructed as in \cite{Shannon_2025}, the true latent population mean, denoted by $\mu_{true,j,t}$, is calculated as the average of the simulated population units within county $j$ in year $t$. From the simulated population, $R = 100$ samples are randomly taken at the same intensity as the observed FIA plot data. Then, for each replicate $r = 1, \ldots, R$, estimates from both the full model and direct estimates are obtained for each county $j$ and time $t$.  Measures of bias, root mean squared error (RMSE), empirical confidence and credible interval coverage rates, and width between the lower and upper 95\% confidence and credible interval bounds, are then used to compare estimators, and are calculated as below.

For estimator $l \in (\text{Direct, Full Model})$, systematic difference between estimate and truth is evaluated using
\begin{equation}
    \text{Bias}_{j,t,l} = \frac{\sum^{R}_{r = 1} \left(\mu_{j,t,l,r} - \mu_{\text{true},j,t}\right)}{R},
\label{eq:bias}
\end{equation}
where $\mu_{j,t,l,r}$ is the $l^{\text{th}}$ estimator's point estimate given sample data from the $r^\text{th}$ replicate. The point estimate for the full model is the posterior mean of $\mu_{j,t}$ and $\hat{\mu}_{j,t}$ for the design-based estimator.

Average accuracy is evaluated using   
\begin{equation}
    \text{RMSE}_{j,t,l} = \sqrt{\frac{\sum^{R}_{r = 1} \left(\mu_{j,t,l,r} - \mu_{\text{true},j,t}\right)^2}{R}}.
\label{eq:rmse}
\end{equation}

Average empirical coverage rate is evaluated using 
\begin{equation}
    \text{Cover}_{j,t,l} = \frac{\sum^{R}_{r = 1} I\left(\mu^L_{j,t,l,r} \le \mu_{\text{true},j,t} \le \mu^U_{j,t,l,r}\right)}{R},
    \label{eq:cover}
\end{equation}
where $I(\cdot)$ is an indicator function, and $\mu^L_{j,t,l,r}$ and $\mu^U_{j,t,l,r}$ are the lower and upper uncertainty quantification bounds. Here, we consider the 95\% credible interval bounds for the full model and 95\% confidence interval bounds for the design-based estimator.

Average precision is evaluated using 
\begin{equation}
    \text{Width}_{j,t,l} = \frac{\sum^{R}_{r = 1} \left(\mu^U_{j,t,l,r} - \mu^L_{j,t,l,r}\right)}{R}.
    \label{eq:width}
\end{equation}

These metrics are compared for both the full model and direct estimator, and are summarized in Figure~\ref{fig:sim_results}.

\clearpage
\section{MCMC Sampler}\label{sec:sampler_computing}

\subsection{Joint Notation}

Following the notation in Section~\ref{sec:model}, the proposed model for county $j$ in year $t$ is

\begin{equation*}
y_{i,j,t} = \underbrace{\bx_{j,t}^\T \bbeta_t + \tilde{\bx}_{j,t}^\T \bet_j + u_{j,t}}_{\mu_{j,t}} + \varepsilon_{i,j,t}, \quad i = i, \ldots, n_{j,t}.
\end{equation*}

Specifications for $\bbeta_t$, $\bet_j$, and $u_{j,t}$, along with appropriate priors as detailed in Section~\ref{sec:model} yield a posterior distribution proportional to the likelihood times priors, given as

\begin{align*}
\begin{split}
&\prod^{T}_{t=1}\prod^J_{j=1} \prod^{n_{j,t}}_{i = 1} N\left(y_{i,j,t}\given \bx_{j,t}^\T \bbeta_t + \tilde{\bx}_{j,t}^\T \bet_{j} + u_{j,t},\, \sigma^2_{t}\right) \times \prod^{T}_{t=1} IG \left( \sigma^2_{t} \given a_\sigma, b_\sigma \right) \times\\
&\quad MVN \left( \bbeta_0 \given \bmu_0, \bSigma_0 \right) \times \prod_{t = 1}^T MVN\left(\bbeta_t \given \bbeta_{t-1}, \bSigma_\xi \right) \times IW \left( \bSigma_\xi \given \nu_\xi, \bH_\xi\right) \times\\
&\quad \prod^{Q}_{q=1} MVN\left(\bet^*_{q} \given \bzero, \tau^2_{\eta, q} \bQ(\rho_{\eta, q}) \right) \times \prod^{Q}_{q=1}  IG \left(\tau^2_{\eta, q} \given a_{\eta,q}, b_{\eta,q} \right) \times \prod^{Q}_{q=1} U\left(\rho_{\eta, q} \given 0, 1\right) \times\\
&\quad \prod^T_{t=1} N\left(\bu_{t} \given \bu_{t-1},\, \tau^2_{\omega, t} \bQ(\rho_{\omega})\right) \times \prod^{T}_{t=1} IG \left(\tau^2_{\omega, t} \given a_{\omega,t}, b_{\omega,t} \right) \times U\left(\rho_{\omega} \given 0, 1 \right).
\end{split}
\end{align*}

To ease computation, we adopt the following stacked notation for model elements. Specifically, let $\by_{t} = (y_{1,1,t}, \ldots, y_{n_{1,t},1,t}, \ldots, y_{1,J,t}, \ldots, y_{n_{J,t},J,t})^\T$ be the length $N_t = \sum_{j=1}^J n_{j,t}$ vector of plot-level measurements observed in time $t$. Then, let $\bX_{t}$ be the $N_t \times (P+1)$ matrix of covariates, whose row entries consist of repeated vectors $\bx_{j,t}$ corresponding to the stacked ordering of $\by_t$. Similarly, let $\tilde{\bX}_{q,t}$ be the $N_t \times J$ block diagonal matrix with $j^\text{th}$ diagonal block being the length $n_{j,t}$ column vector whose elements repeat the $q^\text{th}$ element of $\tilde{\bx}_{j,t}$. Additionally, let $\bet^*_q$ be as defined in Section~\ref{sec:car}. Finally, let $\bu_t$ be defined as in Section~\ref{sec:dynamic}, and $\bA_t$ be the $N_t \times J$ matrix mapping elements of $\bu_t$ to $\by_t$. Then, the proposed model may be written jointly for time $t$ as

\begin{equation*}
\by_{t} = \bX_{t} \bbeta_t + \sum_{q = 1}^Q \tilde{\bX}_{q,t} \bet^*_{q} + \bA_t \bu_{t} + \bvarepsilon_{t},
\end{equation*}

where $\bvarepsilon_t$ is a length $N_t$ vector following a mean zero multivariate normal distribution with covariance matrix $\bSigma_t = \sigma^2_t \bI$, and $\bI$ is the identity matrix. This joint specification allows us to rewrite the proportional posterior above as

\begin{align}\label{eq:joint_post}
\begin{split}
&\prod^{T}_{t=1} MVN\left(\by_{t}\given \bX_{t} \bbeta_t + \sum_{q = 1}^Q \tilde{\bX}_{q,t} \bet^*_{q} + \bA_t \bu_{t},\, \bSigma_{t}\right) \times \prod^{T}_{t=1} IG \left( \sigma^2_{t} \given a_\sigma, b_\sigma \right) \times\\
&\quad MVN \left( \bbeta_0 \given \bmu_0, \bSigma_0 \right) \times \prod_{t = 1}^T MVN\left(\bbeta_t \given \bbeta_{t-1}, \bSigma_\xi \right) \times IW \left( \bSigma_\xi \given \nu_\xi, \bH_\xi\right) \times\\
&\quad \prod^{Q}_{q=1} MVN\left(\bet^*_{q} \given \bzero, \tau^2_{\eta, q} \bQ(\rho_{\eta, q}) \right) \times \prod^{Q}_{q=1}  IG \left(\tau^2_{\eta, q} \given a_{\eta,q}, b_{\eta,q} \right) \times \prod^{Q}_{q=1} U\left(\rho_{\eta, q} \given 0, 1\right) \times\\
&\quad \prod^T_{t=1} MVN\left(\bu_{t} \given \bu_{t-1},\, \tau^2_{\omega, t} \bQ(\rho_{\omega})\right) \times \prod^{T}_{t=1} IG \left(\tau^2_{\omega, t} \given a_{\omega,t}, b_{\omega,t} \right) \times U\left(\rho_{\omega} \given 0, 1 \right).
\end{split}
\end{align}

This joint specification allows for full conditionals to take the form of multivariate normal distributions for many model parameters, which eases notation moving forward.

We now outline the steps to update each model parameter in time order (for $t = 1, \ldots, T$). In cases of conjugacy, full conditional distributional forms will be readily available, and derivations of full conditional distributional parameters will be detailed. For parameters whose prior distributions do not yield conjugate forms, a Metropolis algorithm will be implemented. 

\subsection{Incompleting the Square}\label{sec:incomp_the_sq}

In the case of normal-normal conjugacy (normal prior with normal likelihood), we know the posterior distribution will also follow a normal distribution. The same is true of multivariate normal distributions. To identify the parameters of a full conditional distribution resulting from the product of multivariate normals, we use the method of ``\textit{incompleting the square}'' (see Section 4.2 in \cite{clark_2007_book}). 

Specifically, we start with the known density of the multivariate normal distribution, which takes the form

\begin{equation*}
p \left(\bx \mid \bmu, \bSigma \right) \propto \text{exp} \left(-\frac{1}{2} (\bx - \bmu)^\T \bSigma^{-1} (\bx - \bmu)\right),
\end{equation*}

where $\propto$ denotes ``proportional to.'' We next focus on the terms inside the exponential, ignoring the $-\frac{1}{2}$ constant. Expanding this term yields

\begin{align}\label{eq:expand_mvn}
(\bx - \bmu)^\T \bSigma^{-1} (\bx - \bmu) &= \bx^\T \bSigma^{-1} \bx - \bx^\T \bSigma^{-1} \bmu - \bmu^\T \bSigma^{-1} \bx + \bmu^\T \bSigma^{-1} \bmu.
\end{align}

Writing the mean as a function of the covariance matrix, we can define 

\begin{align*}
\bmu &= \bV \bv, \\
\bSigma &= \bV.
\end{align*}

This allows us to rewrite (\ref{eq:expand_mvn}) as

\begin{align*}
\bx^\T \bV^{-1} \bx &- \bx^\T \bV^{-1} \bV \bv - \bv^\T \bV^\T \bV^{-1} \bx + \bv^\T \bV^\T \bV^{-1} \bV \bv \\
&= \bx^\T \bV^{-1} \bx - \bx^\T \bv - \bv^\T \bx + \bv^\T \bV \bv \\
&= \bx^\T \bV^{-1} \bx - \bx^\T \bv \cdots.
\end{align*}

This expansion allows us to recover $\bV^{-1}$ in the first term $\bx^\T \bV^{-1} \bx$, and $\bv$ in the second term $\bx^\T \bv$. Therefore, when we multiply multivariate normal densities, we can simply combine like terms and identify $\bV^{-1}$ and $\bv$ given their positions in the expanded exponential component. The full conditional distribution will then be $MVN(\bV \bv, \bV)$. 

\subsection{Sampling Steps}

Here, we outline the steps to update each of the model parameters, where text colors for different terms are varied to aid in understanding how the different components of (\ref{eq:joint_post}) contribute to full conditional distributions and log-target densities. 

\subsubsection{Update $\bbeta_0$}

Viewing (\ref{eq:joint_post}), $\bbeta_0$ appears in two terms, namely in \textcolor{RoyalBlue}{its own prior distribution} and in \textcolor{RubineRed}{the prior distribution of $\bbeta_1$}. The product of these distributions is the product of multivariate normal distributions, taking the form

\begin{equation*}
\textcolor{RoyalBlue}{MVN \left( \bbeta_0 \given \bmu_0, \bSigma_0 \right)} \times \textcolor{RubineRed}{MVN\left(\bbeta_1 \given \bbeta_{0}, \bSigma_\xi \right)}.
\end{equation*}

The product of these densities is proportional to

\begin{align*}
&\textcolor{RoyalBlue}{\text{exp} \left(-\frac{1}{2} (\bbeta_0 - \bmu_0)^\T \bSigma_0^{-1} (\bbeta_0 - \bmu_0)\right)} \times \textcolor{RubineRed}{\text{exp} \left(-\frac{1}{2} (\bbeta_1 - \bbeta_0)^\T \bSigma_\xi^{-1} (\bbeta_1 - \bbeta_0)\right)} \\
&\quad \propto \text{exp} \left[ \textcolor{RoyalBlue}{(\bbeta_0 - \bmu_0)^\T \bSigma_0^{-1} (\bbeta_0 - \bmu_0)} + \textcolor{RubineRed}{(\bbeta_1 - \bbeta_0)^\T \bSigma_\xi^{-1} (\bbeta_1 - \bbeta_0)} \right] \\
&\quad = \text{exp} \left[ \textcolor{RoyalBlue}{\bbeta_0^\T \bSigma_0^{-1} \bbeta_0 - \bbeta_0^\T \bSigma_0^{-1} \bmu_0 - \bmu_0^\T \bSigma_0^{-1} \bbeta_0 + \bmu_0^\T \bSigma_0^{-1} \bmu_0 }\right. \\
&\quad \quad \quad \quad + \left. \textcolor{RubineRed}{\bbeta_1^\T \bSigma_\xi^{-1} \bbeta_1 - \bbeta_1^\T \bSigma_\xi^{-1} \bbeta_0 - \bbeta_0^\T \bSigma_\xi^{-1} \bbeta_1 + \bbeta_0^\T \bSigma_\xi^{-1} \bbeta_0 }\right] \\
& \quad \propto \text{exp} \left[ \bbeta_0^\T \left( \textcolor{RoyalBlue}{\bSigma_0^{-1}} + \textcolor{RubineRed}{\bSigma_\xi^{-1}} \right ) \bbeta_0 - \bbeta_0^\T \left( \textcolor{RoyalBlue}{\bSigma_0^{-1}\bmu_0} + \textcolor{RubineRed}{\bSigma_\xi^{-1} \bbeta_1}\right) \right],
\end{align*}

so we can identify

\begin{align*}
&\bV^{-1} =  \textcolor{RoyalBlue}{\bSigma_0^{-1}} + \textcolor{RubineRed}{\bSigma_\xi^{-1}}, \\
&\bv =  \textcolor{RoyalBlue}{\bSigma_0^{-1}\bmu_0} + \textcolor{RubineRed}{\bSigma_\xi^{-1} \bbeta_1},
\end{align*}

and we can sample $\bbeta_0$ from its full conditional distribution as $MVN \left( \bV \bv, \bV \right)$.

\subsubsection{Update $\bet^*_k$ for $k = 1, \ldots, Q$}

Viewing (\ref{eq:joint_post}), $\bet^*_k$ appears in $(T+1)$ terms, namely in \textcolor{RoyalBlue}{its own prior distribution} and in \textcolor{ForestGreen}{the likelihood of $\by_t, t = 1, \ldots, T$}. In the likelihood, we isolate the term containing $\bet^*_k$. Then, the product of these distributions takes the form

\begin{equation*}
\textcolor{RoyalBlue}{MVN \left( \bet^*_k \given \bzero, \tau^2_{\eta, k} \bQ(\rho_{\eta, k}) \right)} \times \textcolor{ForestGreen}{ \prod_{t = 1}^T MVN\Biggl(\by_{t}\given \bX_{t} \bbeta_t + \sum_{\substack{q=1 \\ q\neq k}}^Q \tilde{\bX}_{q,t} \bet^*_{q} + \tilde{\bX}_{k,t} \bet^*_k + \bA_t \bu_{t},\, \bSigma_{t} \Biggl)}.
\end{equation*}

The product of these densities is proportional to

\begin{align*}
&\quad \quad \textcolor{RoyalBlue}{\text{exp} \left(\bet^{*\T}_k (\tau^2_{\eta, k} \bQ(\rho_{\eta, k}))^{-1} \bet^*_k\right)} \times \textcolor{ForestGreen}{\prod_{t = 1}^T \text{exp} \Biggr[ \Biggl(\by_t - \bX_{t} \bbeta_t - \sum_{\substack{q=1 \\ q\neq k}}^Q \tilde{\bX}_{q,t} \bet^*_{q} - \tilde{\bX}_{k,t} \bet^*_k - \bA_t \bu_{t} \Biggl)^\T \Biggr.}\\
&\quad \quad \quad \quad \quad \quad \quad \quad \quad \quad \quad \quad \quad \quad \quad \quad \quad \Biggr. \textcolor{ForestGreen}{\bSigma_t^{-1} \Biggl( \by_t - \bX_{t} \bbeta_t - \sum_{\substack{q=1 \\ q\neq k}}^Q \tilde{\bX}_{q,t} \bet^*_{q} - \tilde{\bX}_{k,t} \bet^*_k - \bA_t \bu_{t}\Biggl) \Biggr]} \\
&= \text{exp} \Biggr[ \textcolor{RoyalBlue}{\bet^{*\T}_k (\tau^2_{\eta, k} \bQ(\rho_{\eta, k}))^{-1} \bet^*_k} + \textcolor{ForestGreen}{\sum_{t = 1}^T \Biggl(\by_t - \bX_{t} \bbeta_t - \sum_{\substack{q=1 \\ q\neq k}}^Q \tilde{\bX}_{q,t} \bet^*_{q} - \tilde{\bX}_{k,t} \bet^*_k - \bA_t \bu_{t} \Biggl)^\T \Biggr.} \\ &\quad \quad \quad \quad \quad \quad \quad \quad \quad \quad \quad \quad \quad \quad \quad \textcolor{ForestGreen}{\Biggr. \bSigma_t^{-1} \Biggl( \by_t - \bX_{t} \bbeta_t - \sum_{\substack{q=1 \\ q\neq k}}^Q \tilde{\bX}_{q,t} \bet^*_{q} - \tilde{\bX}_{k,t} \bet^*_k - \bA_t \bu_{t}\Biggl)}\Biggr] \\
&\propto \text{exp} \Biggr[ \textcolor{RoyalBlue}{\bet^{*\T}_k (\tau^2_{\eta, k} \bQ(\rho_{\eta, k}))^{-1} \bet^*_k} \Biggr. \\
&\quad \quad \Biggr. + \textcolor{ForestGreen}{\sum_{t = 1}^T \Biggr[ \bet^{*\T}_k \tilde{\bX}_{k,t}^\T \bSigma_t^{-1} \tilde{\bX}_{k,t} \bet^*_k - \bet^{*\T}_k \tilde{\bX}_{k,t}^\T \bSigma_t^{-1} \Biggl( \by_t - \bX_{t} \bbeta_t - \sum_{\substack{q=1 \\ q\neq k}}^Q \tilde{\bX}_{q,t} \bet^*_{q} - \bA_t \bu_{t} \Biggl) \Biggr] }\Biggr] \\
&= \text{exp} \Biggr[ \bet^{*\T}_k \left(\textcolor{RoyalBlue}{(\tau^2_{\eta, k} \bQ(\rho_{\eta, k}))^{-1}} + \textcolor{ForestGreen}{\sum_{t = 1}^T \tilde{\bX}_{k,t}^\T \bSigma_t^{-1} \tilde{\bX}_{k,t}} \right) \bet^*_k \Biggr. \\
&\quad \quad \Biggr. - \bet^{*\T}_k \Biggl( \textcolor{ForestGreen}{\sum_{t = 1}^T \tilde{\bX}_{k,t}^\T \bSigma_t^{-1} \Biggl( \by_t - \bX_{t} \bbeta_t - \sum_{\substack{q=1 \\ q\neq k}}^Q \tilde{\bX}_{q,t} \bet^*_{q} - \bA_t \bu_{t} \Biggl)}\Biggl) \Biggr]
\end{align*}

so we can identify

\begin{align*}
&\bV^{-1} =  \textcolor{RoyalBlue}{(\tau^2_{\eta, k} \bQ(\rho_{\eta, k}))^{-1}} + \textcolor{ForestGreen}{\sum_{t = 1}^T \tilde{\bX}_{k,t}^\T \bSigma_t^{-1} \tilde{\bX}_{k,t}} , \\
&\bv = \textcolor{ForestGreen}{\sum_{t = 1}^T \tilde{\bX}_{k,t}^\T \bSigma_t^{-1} \Biggl( \by_t - \bX_{t} \bbeta_t - \sum_{\substack{q=1 \\ q\neq k}}^Q \tilde{\bX}_{q,t} \bet^*_{q} - \bA_t \bu_{t} \Biggl)},
\end{align*}

and we can sample $\bet^*_k$ from its full conditional distribution as $MVN \left( \bV \bv, \bV \right)$.

\subsubsection{Update $\tau^2_{\eta,k}$ for $k = 1, \ldots, Q$}

Viewing (\ref{eq:joint_post}), $\tau^2_{\eta,k}$ appears in two terms, namely in \textcolor{RubineRed}{its own prior distribution} and in \textcolor{RoyalBlue}{the prior for $\bet^*_k$}. The product of these distributions takes the form

\begin{equation*}
\textcolor{RubineRed}{IG \left( \tau^2_{\eta,k} \given a_{\eta,k}, b_{\eta,k} \right)} \times \textcolor{RoyalBlue}{MVN \left( \bet^*_k \given \bzero, \tau^2_{\eta, k} \bQ(\rho_{\eta, k}) \right)}.
\end{equation*}

The product of these densities is proportional to

\begin{align*}
&\textcolor{RubineRed}{\frac{b_{\eta,k}^{a_{\eta,k}}}{\Gamma(a_{\eta,k})} (\tau^2_{\eta,k})^{-a_{\eta,k} - 1} \text{exp} \left( \frac{-b_{\eta,k}}{\tau^2_{\eta,k}} \right)} \times \textcolor{RoyalBlue}{(2\pi)^{-\frac{J}{2}} |\tau^2_{\eta, k} \bQ(\rho_{\eta, k})|^{-\frac{1}{2}} \text{exp} \left( -\frac{1}{2} \bet^{*\T}_k (\tau^2_{\eta, k} \bQ(\rho_{\eta, k}))^{-1} \bet^*_k \right)} \\
&\quad \quad \propto (\tau^2_{\eta, k})^{\textcolor{RoyalBlue}{-\frac{J}{2}} \textcolor{RubineRed}{-a_{\eta,k} - 1}} \text{exp}\left( \frac{\textcolor{RoyalBlue}{-\frac{1}{2} \bet^{*\T}_k \bQ(\rho_{\eta, k})^{-1} \bet^*_k} \textcolor{RubineRed}{- b_{\eta,k}}}{\tau^2_{\eta,k}}\right),
\end{align*}

which is the form of an Inverse Gamma density with shape parameter equal to $\textcolor{RubineRed}{a_{\eta,k}} + \textcolor{RoyalBlue}{\frac{J}{2}}$ and scale parameter $\textcolor{RubineRed}{b_{\eta,k}} + \textcolor{RoyalBlue}{\frac{1}{2} \bet^{*\T}_k \bQ(\rho_{\eta, k})^{-1} \bet^*_k}$.

So, we sample $\tau^2_{\eta,k}$ from $IG \left(\textcolor{RubineRed}{a_{\eta,k}} + \textcolor{RoyalBlue}{\frac{J}{2}}, \textcolor{RubineRed}{b_{\eta,k}} + \textcolor{RoyalBlue}{\frac{1}{2} \bet^{*\T}_k \bQ(\rho_{\eta, k})^{-1} \bet^*_k}\right)$. 

\subsubsection{Update $\bSigma_{\xi}$}

Viewing (\ref{eq:joint_post}), $\bSigma_\xi$ appears in $T+1$ terms, namely in \textcolor{RubineRed}{its own prior distribution}, and in \textcolor{RoyalBlue}{the prior for $\bbeta_t, t = 1, \ldots, T$}. The product of these distributions takes the form

\begin{equation*}
\textcolor{RoyalBlue}{ \prod_{t=1}^T MVN(\bbeta_t \mid \bbeta_{t-1}, \bSigma_\xi)} \times \textcolor{RubineRed}{IW(\bSigma_\xi \mid \nu_\xi, \bH_\xi)}
\end{equation*}

The product of these densities becomes

\begin{align*}
&\textcolor{RoyalBlue}{\prod_{t=1}^T (2\pi)^{-(P+1)/2} |\bSigma_\xi|^{-1/2} \text{exp} \left( -\frac{1}{2} (\bbeta_t - \bbeta_{t-1})^\T \bSigma_\xi^{-1} (\bbeta_t - \bbeta_{t-1}) \right)} \\ 
&\quad \times \textcolor{RubineRed}{\frac{|\bH_\xi|^{\frac{\nu_\xi}{2}}}{2^{\frac{\nu_\xi (P+1)}{2}}  \Gamma_{(P+1)} \left( \frac{\nu_\xi}{2} \right)}  |\bSigma_\xi|^{-\frac{\nu_\xi + P + 2}{2}} \text{exp}\left(-\frac{1}{2} \text{tr}\left( \bH_\xi \bSigma_\xi^{-1} \right) \right)} \\
&\propto |\bSigma_\eta|^{- \frac{\textcolor{RoyalBlue}{T} + \textcolor{RubineRed}{v+P+2}}{2}} \text{exp} \left( \frac{-1}{2} \left[ \underbrace{\textcolor{RoyalBlue}{\sum_{t = 1}^T (\bbeta_t - \bbeta_{t-1})^\T \bSigma_\xi^{-1} (\bbeta_t - \bbeta_{t-1})}}_A + \textcolor{RubineRed}{\text{tr} ( \bH_\xi \bSigma_\xi^{-1} )} \right] \right)
\end{align*}

Since $A$ is evaluated as a scalar (sum of quadratic forms), we can take the \textit{trace} of it without changing the expression (the trace of a scalar is the scalar itself). So we can write

\begin{align*}
\quad A &= \textcolor{RoyalBlue}{\sum_{t = 1}^T (\bbeta_t - \bbeta_{t-1})^\T \bSigma_\xi^{-1} (\bbeta_t - \bbeta_{t-1})} \\
\quad &= \text{tr} \left( \textcolor{RoyalBlue}{\sum_{t = 1}^T (\bbeta_t - \bbeta_{t-1})^\T \bSigma_\xi^{-1} (\bbeta_t - \bbeta_{t-1})} \right) \\
\quad &=  \textcolor{RoyalBlue}{\sum_{t = 1}^T} \text{tr} \left( \textcolor{RoyalBlue}{(\bbeta_t - \bbeta_{t-1})^\T \bSigma_\xi^{-1} (\bbeta_t - \bbeta_{t-1})} \right) \\
\quad &=  \textcolor{RoyalBlue}{\sum_{t = 1}^T} \text{tr} \left( \textcolor{RoyalBlue}{(\bbeta_t - \bbeta_{t-1}) (\bbeta_t - \bbeta_{t-1})^\T \bSigma_\xi^{-1}} \right),
\end{align*}

and plugging $A$ back in gives

\begin{align*}
&|\bSigma_\eta|^{- \frac{\textcolor{RoyalBlue}{T} + \textcolor{RubineRed}{v+P+2}}{2}} \text{exp} \left( \frac{-1}{2} \left[ \textcolor{RoyalBlue}{\sum_{t = 1}^T (\bbeta_t - \bbeta_{t-1})^\T \bSigma_\xi^{-1} (\bbeta_t - \bbeta_{t-1})} + \textcolor{RubineRed}{\text{tr} ( \bH_\xi \bSigma_\xi^{-1} )} \right] \right) \\
= &|\bSigma_\eta|^{- \frac{\textcolor{RoyalBlue}{T} + \textcolor{RubineRed}{v+P+2}}{2}} \text{exp} \left( -\frac{1}{2} \text{tr} \left[ \left( \textcolor{RoyalBlue}{\sum_{t = 1}^T (\bbeta_t - \bbeta_{t-1}) (\bbeta_t - \bbeta_{t-1})^\T} + \textcolor{RubineRed}{\bH_\xi} \right) \bSigma_\xi^{-1} \right] \right)
\end{align*}

which is in the same form of an inverse Wishart density with parameter values

\begin{align*}
\tilde{\nu}_\xi &= \textcolor{RubineRed}{\nu_\xi} + \textcolor{RoyalBlue}{T} \\
\tilde{\bH}_\xi &= \textcolor{RubineRed}{\bH_\xi} + \textcolor{RoyalBlue}{\sum_{t = 1}^T (\bbeta_t - \bbeta_{t-1}) (\bbeta_t - \bbeta_{t-1})^\T}
\end{align*}

and we can update $\bSigma_\xi$ from its full conditional distribution as $IW(\tilde{\bH}_\xi, \tilde{\nu}_\xi)$.

\subsubsection{Update $\rho_{\eta, q}$ for $q = 1, \ldots, Q$}

Viewing (\ref{eq:joint_post}), $\rho_{\eta,q}$ appears in two terms, namely in \textcolor{RubineRed}{its own prior distribution} and in \textcolor{RoyalBlue}{the prior for $\bet^*_q$}. The product of these densities is

\begin{align*}
&\textcolor{RoyalBlue}{MVN(\bet^*_k \mid \bzero, \tau^2_{\eta,q} \bQ(\rho_{\eta,q}))} \times \textcolor{RubineRed}{U(\rho_{\eta,q} \mid a_\eta, b_\eta)} \\
&\quad \quad = \textcolor{RoyalBlue}{(2 \pi)^{-\frac{J}{2}} |\tau^2_{\eta,q} \bQ(\rho_{\eta,q})|^{-\frac{1}{2}} \text{exp} \left( - \frac{1}{2} \bet^{*\T}_q (\tau^2_{\eta,q} \bQ(\rho_{\eta,q}))^{-1} \bet^*_q \right)}  \times \textcolor{RubineRed}{\frac{1}{b_\eta - a_\eta} \boldsymbol{1}_{a_\eta < \rho_{\eta,q} < b_\eta}}
\end{align*}

where $a_\eta = 0$, $b_\eta = 1$, and $\boldsymbol{1}$ is the indicator function. Next, we employ a transformation on the Uniform variable $\rho_{\eta,q}$ of the form

\begin{equation*}
g(\rho_{\eta,q}) = \log \left( \frac{\rho_{\eta,q} - a_\eta}{b_\eta - \rho_{\eta,q}} \right).
\end{equation*}

To get the distribution of this newly transformed variable $g(\rho_{\eta,q})$, we require the Jacobian. First, we find the inverse function of the transformation, which is 

\begin{equation*}
g^{-1}(g(\rho_{\eta,q})) = \frac{b_\eta e^{g(\rho_{\eta,q})} + a_\eta}{e^{g(\rho_{\eta,q})} + 1}.
\end{equation*}

Next, we take the derivative of this function with respect to $g(\rho_{\eta,q})$, which gives

\begin{equation*}
\frac{e^{g(\rho_{\eta,q})} (b_\eta - a_\eta)}{(e^{g(\rho_{\eta,q})} + 1)^2},
\end{equation*}

and plugging the expression for $g(\rho_{\eta,q})$ into this equation yields

\begin{equation*}
\frac{\frac{\rho_{\eta,q} - a_\eta}{b_\eta - \rho_{\eta,q}} (b_\eta - a_\eta)}{\left(\frac{\rho_{\eta,q} - a_\eta}{b_\eta - \rho_{\eta,q}} + 1 \right)^2} \propto \frac{\frac{\rho_{\eta,q} - a_\eta}{b_\eta - \rho_{\eta,q}}}{\left(\frac{b_\eta - \rho_{\eta,q} + \rho_{\eta, q} - a_\eta}{b_\eta - \rho_{\eta,q}} \right)^2} \propto \frac{\frac{\rho_{\eta,q} - a_\eta}{b_\eta - \rho_{\eta,q}}}{\frac{1}{(b_\eta - \rho_{\eta,q})^2}} = (\rho_{\eta,q} - a_\eta)(b_\eta - \rho_{\eta,q}).
\end{equation*}

With this jacobian adjustment, we can now determine the \textit{log target density} to which we will add the Jacobian adjustment. We have

\begin{align*}
&\log \biggl[ \textcolor{RoyalBlue}{(2 \pi)^{-\frac{J}{2}} |\tau^2_{\eta,q} \bQ(\rho_{\eta,q})|^{-\frac{1}{2}} \text{exp} \left( - \frac{1}{2} \bet^{*\T}_q (\tau^2_{\eta,q} \bQ(\rho_{\eta,q}))^{-1} \bet^*_q \right)}  \times \textcolor{RubineRed}{\frac{1}{b_\eta - a_\eta} \boldsymbol{1}_{a_\eta < \rho_{\eta,q} < b_\eta}} \biggr. \\ 
&\quad \quad \biggr. \times (\rho_{\eta,q} - a_\eta)(b_\eta - \rho_{\eta,q}) \biggr] \\
&\quad \quad \propto \textcolor{RoyalBlue}{- \frac{1}{2} \log |\tau_{\eta,q}^2 \bQ(\rho_{\eta,q})| - \frac{1}{2} \bet^{*\T}_q (\tau_{\eta,q}^2 \bQ(\rho_{\eta,q}))^{-1} \bet^*_q} + \log (\rho_{\eta,q} - a_\eta) + \log (b_\eta - \rho_{\eta,q}) 
\end{align*}

We can then proceed by generating a proposed value of $g(\rho_{\eta,q})$ from some proposal distribution (usually Gaussian), evaluating the log target density plus Jacobian adjustment, and then accepting or rejecting the back-transformed $g^{-1}(g(\rho_{\eta,q}))$ based on this acceptance probability. 

\subsubsection{Update $\rho_\omega$}

Viewing (\ref{eq:joint_post}), $\rho_\omega$ appears in $(T+1)$ terms, namely in \textcolor{RoyalBlue}{its own prior distribution} and in \textcolor{RubineRed}{the prior for $\bu_t, t = 1, \ldots, T$}. The product of these densities is

\begin{align*}
&\textcolor{RubineRed}{ \prod_{t = 1}^T MVN(\bu_t \mid \bu_{t-1}, \tau^2_{\omega,t} \bQ(\rho_\omega))} \times \textcolor{RoyalBlue}{U(\rho_\omega \mid a_\omega, b_\omega)} \\
&\quad \quad = \textcolor{RubineRed}{\prod_{t = 1}^T (2 \pi)^{-\frac{J}{2}} |\tau^2_{\omega,t} \bQ(\rho_\omega)|^{-\frac{1}{2}} \text{exp} \left( - \frac{1}{2} (\bu_t - \bu_{t-1})^\T (\tau^2_{\omega,t} \bQ(\rho_\omega))^{-1} (\bu_t - \bu_{t- 1}) \right)}\\
&\quad \quad \quad \quad\times \textcolor{RoyalBlue}{\frac{1}{b_\omega - a_\omega} \boldsymbol{1}_{a_\omega < \rho_\omega < b_\omega}}
\end{align*}

where $a_\omega = 0$, $b_\omega = 1$, and $\boldsymbol{1}$ is the indicator function. We then calculate the log target density to which we will add a Jacobian adjustment. We have

\begin{align*}
&\log \left[ \textcolor{RubineRed}{\prod_{t = 1}^T (2 \pi)^{-\frac{J}{2}} |\tau^2_{\omega,t} \bQ(\rho_\omega)|^{-\frac{1}{2}} \text{exp} \left( - \frac{1}{2} (\bu_t - \bu_{t-1})^\T (\tau^2_{\omega,t} \bQ(\rho_\omega))^{-1} (\bu_t - \bu_{t- 1}) \right)} \right. \\
&\quad \quad \left. \times \textcolor{RoyalBlue}{\frac{1}{b_\omega - a_\omega} \boldsymbol{1}_{a_\omega < \rho_\omega < b_\omega}} \times (\rho_\omega - a_\omega)(b_\omega - \rho_\omega) \right] \\
&\propto \textcolor{RubineRed}{- \frac{1}{2} \sum_{t = 1}^T \Bigl( \log |\tau_{\omega,t}^2 \bQ(\rho_\omega)| + (\bu_t - \bu_{t-1})^\T (\tau_{\omega,t}^2 \bQ(\rho_\omega))^{-1} (\bu_t - \bu_{t-1}) \Bigl)} \\
&\quad \quad + \log (\rho_\omega - a_\omega) + \log (b_\omega - \rho_\omega) 
\end{align*}

We can then proceed by generating a proposed value of $g(\rho_\omega)$ from some proposal distribution (usually Gaussian), evaluating the log target density plus Jacobian adjustment, and then accepting or rejecting the back-transformed $g^{-1}(g(\rho_\omega))$ based on this acceptance probability. 

\subsubsection*{Then for $t = 1, \ldots, T$, do the following:}

\subsubsection{Update $\bbeta_t$}

Viewing (\ref{eq:joint_post}), $\bbeta_t$ appears in three terms, namely in \textcolor{RoyalBlue}{its own prior distribution}, \textcolor{RubineRed}{the prior distribution of $\bbeta_{t+1}$}, and \textcolor{ForestGreen}{the likelihood for $\by_t$}. The product of these distributions takes the form

\begin{align*}
&\textcolor{RoyalBlue}{MVN \left( \bbeta_t \given \bbeta_{t-1}, \bSigma_\xi \right)} \times \textcolor{RubineRed}{MVN\left(\bbeta_{t+1} \given \bbeta_t, \bSigma_\xi \right)} \\
&\times \textcolor{ForestGreen}{MVN \biggl(\by_{t}\given \bX_{t} \bbeta_t + \sum_{q = 1}^Q \tilde{\bX}_{q,t} \bet^*_{q} + \bA_t \bu_{t},\, \bSigma_{t} \biggl)}.
\end{align*}

which is proportional to

\begin{align*}
&\textcolor{RoyalBlue}{\text{exp} \left(-\frac{1}{2} (\bbeta_t - \bbeta_{t-1})^\T \bSigma_\xi^{-1} (\bbeta_t - \bbeta_{t-1})\right)} \times \textcolor{RubineRed}{\text{exp} \left(-\frac{1}{2} (\bbeta_{t+1} - \bbeta_t)^\T \bSigma_\xi^{-1} (\bbeta_{t+1} - \bbeta_t)\right)} \\
&\quad \times \textcolor{ForestGreen}{\text{exp} \biggr[ \biggl( \by_t - \bX_{t} \bbeta_t - \sum_{q = 1}^Q \tilde{\bX}_{q,t} \bet^*_{q} - \bA_t \bu_{t} \biggl)^\T \bSigma_t^{-1} \biggl( \by_t - \bX_{t} \bbeta_t - \sum_{q = 1}^Q \tilde{\bX}_{q,t} \bet^*_{q} - \bA_t \bu_{t} \biggl)\biggr]} \\
&\propto \text{exp} \biggr[ \textcolor{RoyalBlue}{(\bbeta_t - \bbeta_{t-1})^\T \bSigma_\xi^{-1} (\bbeta_t - \bbeta_{t-1})} + \textcolor{RubineRed}{(\bbeta_{t+1} - \bbeta_t)^\T \bSigma_\xi^{-1} (\bbeta_{t+1} - \bbeta_t)} \biggr. \\
&\quad \biggr. + \textcolor{ForestGreen}{\biggl( \by_t - \bX_{t} \bbeta_t - \sum_{q = 1}^Q \tilde{\bX}_{q,t} \bet^*_{q} - \bA_t \bu_{t} \biggl)^\T \bSigma_t^{-1} \biggl( \by_t - \bX_{t} \bbeta_t - \sum_{q = 1}^Q \tilde{\bX}_{q,t} \bet^*_{q} - \bA_t \bu_{t} \biggl)} \biggr] \\
&\propto \text{exp} \biggr[ \textcolor{RoyalBlue}{\bbeta_t^\T \bSigma_\xi^{-1} \bbeta_t - \bbeta_t^\T \bSigma_\xi^{-1} \bbeta_{t-1}} + \textcolor{RubineRed}{\bbeta_t^\T \bSigma_\xi^{-1} \bbeta_t - \bbeta_t^\T \bSigma_\xi^{-1} \bbeta_{t+1}}\biggr. \\
&\quad \quad \quad \quad + \biggr. \textcolor{ForestGreen}{\bbeta_t^\T \bX^\T_{t} \bSigma_t^{-1} \bX_t \bbeta_t - \bbeta_t^\T \bX^\T_t \bSigma_t^{-1} \biggl(\by_t - \sum_{q = 1}^Q \tilde{\bX}_{q,t} \bet^*_{q} - \bA_t \bu_{t}\biggl)}\biggr] \\
&= \text{exp} \biggr[ \bbeta_t^\T \left( \textcolor{RoyalBlue}{\bSigma_\xi^{-1}} + \textcolor{RubineRed}{\bSigma_\xi^{-1}} + \textcolor{ForestGreen}{\bX^\T_{t} \bSigma_t^{-1} \bX_t} \right) \bbeta_t \biggr.\\
&\quad \quad \quad \quad \biggr. - \bbeta_t^\T \biggl( \textcolor{RoyalBlue}{\bSigma_\xi^{-1}\bbeta_{t-1}} + \textcolor{RubineRed}{\bSigma_\xi^{-1} \bbeta_{t+1}} + \textcolor{ForestGreen}{\bX^\T_t \bSigma_t^{-1} \biggl(\by_t - \sum_{q = 1}^Q \tilde{\bX}_{q,t} \bet^*_{q} - \bA_t \bu_{t}\biggl)} \biggl) \biggr],
\end{align*}

so we can identify

\begin{align*}
&\bV^{-1} = \textcolor{RoyalBlue}{\bSigma_\xi^{-1}} + \textcolor{RubineRed}{\bSigma_\xi^{-1}} + \textcolor{ForestGreen}{\bX^\T_{t} \bSigma_t^{-1} \bX_t}, \\
&\bv = \textcolor{RoyalBlue}{\bSigma_\xi^{-1}\bbeta_{t-1}} + \textcolor{RubineRed}{\bSigma_\xi^{-1} \bbeta_{t+1}} + \textcolor{ForestGreen}{\bX^\T_t \bSigma_t^{-1} \biggl(\by_t - \sum_{q = 1}^Q \tilde{\bX}_{q,t} \bet^*_{q} - \bA_t \bu_{t}\biggl)},
\end{align*}

and we can sample $\bbeta_t$ from its full conditional distribution as $MVN \left( \bV \bv, \bV \right)$.

For $t = T$, we instead have

\begin{align*}
&\bV^{-1} = \textcolor{RoyalBlue}{\bSigma_\xi^{-1}} + \textcolor{ForestGreen}{\bX^\T_{t} \bSigma_T^{-1} \bX_t}, \\
&\bv = \textcolor{RoyalBlue}{\bSigma_\xi^{-1}\bbeta_{T-1}} + \textcolor{ForestGreen}{\bX^\T_T \bSigma_T^{-1} \biggl(\by_T - \sum_{q = 1}^Q \tilde{\bX}_{q,T} \bet^*_{q} - \bA_T \bu_{T}\biggl)},
\end{align*}

and we can sample $\bbeta_T$ from its full conditional distribution as $MVN \left( \bV \bv, \bV \right)$.

\subsubsection{Update $\bu_t$}

Viewing (\ref{eq:joint_post}), $\bu_t$ appears in three terms, namely \textcolor{ForestGreen}{the likelihood for $\by_t$}, \textcolor{RoyalBlue}{the prior for $\bu_t$}, and in \textcolor{RubineRed}{the prior for $\bu_{t+1}$}. The product of these distributions takes the form

\begin{align*}
&\textcolor{RoyalBlue}{MVN \left(\bu_t \mid \bu_{t-1}, \tau^2_{\omega,t} \bQ(\rho_\omega)\right)} \times \textcolor{RubineRed}{MVN \left(\bu_{t+1} \mid \bu_{t}, \tau^2_{\omega, t+1} \bQ(\rho_\omega) \right)} \\
&\quad \quad \times \textcolor{ForestGreen}{MVN \biggl(\by_{t}\given \bX_{t} \bbeta_t + \sum_{q = 1}^Q \tilde{\bX}_{q,t} \bet^*_{q} + \bA_t \bu_{t},\, \bSigma_{t}  \biggl)}.
\end{align*}

The product of these densities is proportional to

\begin{align*}
&\textcolor{RoyalBlue}{\text{exp} \left(-\frac{1}{2} (\bu_t - \bu_{t-1})^\T (\tau^2_{\omega,t} \bQ(\rho_\omega))^{-1} (\bu_t - \bu_{t-1})\right)} \\
&\quad \quad \times \textcolor{RubineRed}{\text{exp} \left(-\frac{1}{2} (\bu_{t+1} - \bu_t)^\T (\tau^2_{\omega,t+1} \bQ(\rho_\omega))^{-1} (\bu_{t+1} - \bu_t)\right)} \\
&\quad \quad \times \textcolor{ForestGreen}{\text{exp} \biggr[ \biggl( \by_t - \bX_{t} \bbeta_t - \sum_{q = 1}^Q \tilde{\bX}_{q,t} \bet^*_{q} - \bA_t \bu_{t} \biggl)^\T \bSigma_t^{-1} \biggl( \by_t - \bX_{t} \bbeta_t - \sum_{q = 1}^Q \tilde{\bX}_{q,t} \bet^*_{q} - \bA_t \bu_{t} \biggl)\biggr]} \\
&\propto \text{exp} \biggr[ \textcolor{RoyalBlue}{(\bu_t - \bu_{t-1})^\T (\tau^2_{\omega,t} \bQ(\rho_\omega))^{-1} (\bu_t - \bu_{t-1})} \\
&\quad \quad + \textcolor{RubineRed}{(\bu_{t+1} - \bu_t)^\T (\tau^2_{\omega,t+1} \bQ(\rho_\omega))^{-1} (\bu_{t+1} - \bu_t)} \biggr. \\
&\quad \quad \biggr. + \textcolor{ForestGreen}{\biggl( \by_t - \bX_{t} \bbeta_t - \sum_{q = 1}^Q \tilde{\bX}_{q,t} \bet^*_{q} - \bA_t \bu_{t} \biggl)^\T \bSigma_t^{-1} \biggl( \by_t - \bX_{t} \bbeta_t - \sum_{q = 1}^Q \tilde{\bX}_{q,t} \bet^*_{q} - \bA_t \bu_{t} \biggl)} \biggr] \\
&\propto \text{exp} \biggr[ \textcolor{RoyalBlue}{\bu_t^\T (\tau^2_{\omega,t} \bQ(\rho_\omega))^{-1} \bu_t - \bu_t^\T (\tau^2_{\omega,t} \bQ(\rho_\omega))^{-1} \bu_{t-1}} \biggr. \\
&\quad \quad + \biggr. \textcolor{RubineRed}{\bu_t^\T (\tau^2_{\omega,t+1} \bQ(\rho_\omega))^{-1} \bu_t - \bu_t^\T (\tau^2_{\omega,t+1} \bQ(\rho_\omega))^{-1} \bu_{t+1}} \biggr. \\
&\quad \quad + \biggr. \textcolor{ForestGreen}{\bu_t^\T \bA_t^\T \bSigma_t^{-1} \bA_t \bu_t - \bu_t^\T \bA^\T_t \bSigma_t^{-1} \biggl(\by_t - \bX_t \bbeta_t - \sum_{q = 1}^Q \tilde{\bX}_{q,t} \bet^*_{q} \biggl)}\biggr] \\
&= \text{exp} \biggr[ \bu_t^\T \left( \textcolor{RoyalBlue}{(\tau^2_{\omega,t} \bQ(\rho_\omega))^{-1}} + \textcolor{RubineRed}{(\tau^2_{\omega,t+1} \bQ(\rho_\omega))^{-1}} + \textcolor{ForestGreen}{\bA^\T_{t} \bSigma_t^{-1} \bA_t} \right) \bu_t \biggr.\\
&\quad \quad \biggr. - \bu_t^\T \biggl( \textcolor{RoyalBlue}{(\tau^2_{\omega,t} \bQ(\rho_\omega))^{-1}\bu_{t-1}} + \textcolor{RubineRed}{(\tau^2_{\omega,t+1} \bQ(\rho_\omega))^{-1} \bu_{t+1}} + \textcolor{ForestGreen}{\bA^\T_t \bSigma_t^{-1} \biggl(\by_t - \bX_t \bbeta_t - \sum_{q = 1}^Q \tilde{\bX}_{q,t} \bet^*_{q} \biggl)} \biggl) \biggr],
\end{align*}

so we can identify 

\begin{align*}
&\bV^{-1} =  \textcolor{RoyalBlue}{(\tau^2_{\omega,t} \bQ(\rho_\omega))^{-1}} + \textcolor{RubineRed}{(\tau^2_{\omega,t+1} \bQ(\rho_\omega))^{-1}} + \textcolor{ForestGreen}{\bA^\T_{t} \bSigma_t^{-1} \bA_t}, \\
&\bv = \textcolor{RoyalBlue}{(\tau^2_{\omega,t} \bQ(\rho_\omega))^{-1}\bu_{t-1}} + \textcolor{RubineRed}{(\tau^2_{\omega,t+1} \bQ(\rho_\omega))^{-1} \bu_{t+1}} + \textcolor{ForestGreen}{\bA^\T_t \bSigma_t^{-1} \biggl(\by_t - \bX_t \bbeta_t - \sum_{q = 1}^Q \tilde{\bX}_{q,t} \bet^*_{q} \biggl)},
\end{align*}

and we can update $\bu_t$ from its full conditional as $MVN(\bV \bv, \bV)$.

For $t = T$, we instead have 

\begin{align*}
&\bV^{-1} =  \textcolor{RoyalBlue}{(\tau^2_{\omega,T} \bQ(\rho_\omega))^{-1}} +  \textcolor{ForestGreen}{\bA^\T_{T} \bSigma_T^{-1} \bA_T}, \\
&\bv = \textcolor{RoyalBlue}{(\tau^2_{\omega,T} \bQ(\rho_\omega))^{-1}\bu_{T-1}} + \textcolor{ForestGreen}{\bA^\T_T \bSigma_T^{-1} \biggl(\by_T - \bX_T \bbeta_T - \sum_{q = 1}^Q \tilde{\bX}_{q,T} \bet^*_{q} \biggl)},
\end{align*}

and we can update $\bu_T$ from its full conditional as $MVN(\bV \bv, \bV)$.

\subsubsection{Update $\tau^2_{\omega,t}$}

Viewing (\ref{eq:joint_post}), $\tau^2_{\omega,t}$ appears in two terms, namely in \textcolor{RoyalBlue}{its own prior distribution} and in \textcolor{RubineRed}{the prior for $\bu_t$}. The product of these distributions takes the form

\begin{equation*}
\textcolor{RoyalBlue}{IG \left( \tau^2_{\omega,t} \given a_{\omega,t}, b_{\omega,t} \right)} \times \textcolor{RubineRed}{MVN \left( \bu_t \given \bu_{t-1}, \tau^2_{\omega, t} \bQ(\rho_{\omega}) \right)},
\end{equation*}

which is proportional to

\begin{align*}
&\textcolor{RoyalBlue}{\frac{b_{\omega,t}^{a_{\omega,t}}}{\Gamma(a_{\omega,t})} (\tau^2_{\omega,t})^{-a_{\omega,t} - 1} \text{exp} \left( \frac{-b_{\omega,t}}{\tau^2_{\omega,t}} \right)} \\
&\quad \quad \times \textcolor{RubineRed}{(2\pi)^{-\frac{J}{2}} |\tau^2_{\omega, t} \bQ(\rho_{\omega})|^{-\frac{1}{2}} \text{exp} \left( -\frac{1}{2} (\bu_t - \bu_{t-1})^\T (\tau^2_{\omega, t} \bQ(\rho_{\omega}))^{-1} (\bu_t - \bu_{t-1}) \right)} \\
&\quad \quad \propto (\tau^2_{\omega, t})^{\textcolor{RubineRed}{-\frac{J}{2}} \textcolor{RoyalBlue}{-a_{\omega,t} - 1}} \text{exp}\left( \frac{\textcolor{RubineRed}{-\frac{1}{2} (\bu_t - \bu_{t-1})^\T \bQ(\rho_{\omega})^{-1} (\bu_t - \bu_{t-1})} \textcolor{RoyalBlue}{- b_{\omega,t}}}{\tau^2_{\omega,t}}\right),
\end{align*}

which is the form of an Inverse Gamma density with shape parameter equal to $\textcolor{RoyalBlue}{a_{\omega,t}} + \textcolor{RubineRed}{\frac{J}{2}}$ and scale parameter $\textcolor{RoyalBlue}{b_{\omega,t}} + \textcolor{RubineRed}{\frac{1}{2} (\bu_t - \bu_{t-1})^\T \bQ(\rho_{\omega})^{-1} (\bu_t - \bu_{t-1})}$.

So, we sample $\tau^2_{\omega,t}$ from $IG \left(\textcolor{RoyalBlue}{a_{\omega,t}} + \textcolor{RubineRed}{\frac{J}{2}}, \textcolor{RoyalBlue}{b_{\omega,t}} + \textcolor{RubineRed}{\frac{1}{2} (\bu_t - \bu_{t-1})^\T \bQ(\rho_{\omega})^{-1} (\bu_t - \bu_{t-1})}\right)$.

\subsubsection{Update $\sigma^2_t$}

Viewing (\ref{eq:joint_post}), $\sigma^2_{t}$ appears in two terms, namely in \textcolor{RubineRed}{its own prior distribution} and in \textcolor{ForestGreen}{the likelihood of $\by_t$}. The product of these distributions takes the form

\begin{equation*}
\textcolor{RubineRed}{IG \left( \sigma^2_{t} \given a_{\sigma}, b_{\sigma} \right)} \times \textcolor{ForestGreen}{MVN \biggl(\by_{t}\given \bX_{t} \bbeta_t + \sum_{q = 1}^Q \tilde{\bX}_{q,t} \bet^*_{q} + \bA_t \bu_{t},\, \bSigma_{t}  \biggl)}.
\end{equation*}

The product of these densities is proportional to

\begin{align*}
&\textcolor{RubineRed}{\frac{b_{\sigma}^{a_{\sigma}}}{\Gamma(a_{\sigma})} (\sigma^2_{t})^{-a_{\sigma} - 1} \text{exp} \left( \frac{-b_{\sigma}}{\sigma^2_{t}} \right)} \times \textcolor{ForestGreen}{\text{exp} \Biggr[ \Biggl(\by_t - \bX_{t} \bbeta_t - \sum_{\substack{q=1 \\ q\neq k}}^Q \tilde{\bX}_{q,t} \bet^*_{q} - \tilde{\bX}_{k,t} \bet^*_k - \bA_t \bu_{t} \Biggl)^\T \Biggr.}\\
&\quad \quad \quad \quad \quad \quad \quad \quad \quad \quad \quad \quad \quad \quad \quad \Biggr. \textcolor{ForestGreen}{\bSigma_t^{-1} \Biggl( \by_t - \bX_{t} \bbeta_t - \sum_{\substack{q=1 \\ q\neq k}}^Q \tilde{\bX}_{q,t} \bet^*_{q} - \tilde{\bX}_{k,t} \bet^*_k - \bA_t \bu_{t}\Biggl) \Biggr]} \\
&\quad \quad \propto (\sigma^2_{t})^{\textcolor{ForestGreen}{-\frac{N_t}{2}} \textcolor{RubineRed}{-a_{\sigma} - 1}} \times \text{exp}\left( \frac{\textcolor{ForestGreen}{-\frac{1}{2} (\by_t - \bmu_t )^\T (\by_t - \bmu_t )} \textcolor{RubineRed}{- b_{\sigma}}}{\sigma^2_{t}}\right),
\end{align*}

where $\bmu_t = \bX_{t} \bbeta_t + \sum_{\substack{q=1 \\ q\neq k}}^Q \tilde{\bX}_{q,t} \bet^*_{q} + \tilde{\bX}_{k,t} \bet^*_k + \bA_t \bu_{t}$.

This gives the form of an Inverse Gamma density with shape parameter equal to $\textcolor{RubineRed}{a_{\sigma}} + \textcolor{ForestGreen}{\frac{N_t}{2}}$ and scale parameter $\textcolor{RubineRed}{b_{\sigma}} + \textcolor{ForestGreen}{\frac{1}{2} (\by_t - \bmu_t )^\T (\by_t - \bmu_t )}$.

So, we sample $\sigma^2_{t}$ from $IG \left(\textcolor{RubineRed}{a_{\sigma}} + \textcolor{ForestGreen}{\frac{N_t}{2}}, \textcolor{RubineRed}{b_{\sigma}} + \textcolor{ForestGreen}{\frac{1}{2} (\by_t - \bmu_t )^\T (\by_t - \bmu_t )}\right)$.

\subsubsection{Update $\mu_{j,t}$ for $j = 1, \ldots, J$.}

The posterior distribution for $\mu_{j,t}$ is simply $\mu_{j,t} = \bx_{j,t} \bbeta_t + \tilde{\bx}_{j,t} \bet_j + u_{j,t}$. 

\section{Computing Notes}\label{sec:computing}

To efficiently evaluate the CAR precision matrix $\bQ(\rho)^{-1}$, we define $\bQ(\rho)^{-1} = (\bD - \rho \bW) = \bD^{1/2}(\bI - \rho \bD^{-1/2}\bW \bD^{-1/2}) \bD^{1/2}$, where $\bI$ is the $J \times J$ identity matrix, and let $\bD^{-1/2} \bW \bD^{-1/2} = \bP\bLambda \bP^{\T}$ where $\bLambda$ is the diagonal matrix of eigenvalues and the columns of $\bP$ are the eigenvectors of $\bD^{-1/2} \bW \bD^{-1/2}$. This allows $\bQ(\rho)^{-1}$ to be expressed as $\sum_{j=1}^J (1 - \rho\lambda_j)\bv_j \bv_j^{\T} = \sum_{j=1}^J \bv_j \bv_j^{\T} - \rho (\sum_{j=1}^J\lambda_j \bv_j \bv_j^{\T})$, where $\bv_j$ are the columns of $\bD^{1/2}\bP$ and $\lambda_j$ are the diagonal elements of $\bLambda$. Expressing the precision matrix in this way removes the need for costly matrix formation and Cholesky decomposition in each MCMC iteration. Rather, the first term, i.e., $\sum_{j=1}^J \bv_j \bv_j^{\T}$, remains the same across MCMC iterations and the second term, i.e., $\rho (\sum_{j=1}^J\lambda_j \bv_j \bv_j^{\T})$, only varies by a multiplicative constant $\rho$. Further, the determinant of $\bQ(\rho)^{-1}$, which is needed to update correlation parameters $\rho_{\eta, q}$ and $\rho_{\omega}$ via Metropolis steps, is simplified to $\prod^J_{j=1}d_j\left(1 - \rho\lambda_j\right)$, which requires no linear algebra across MCMC iterations. 

Additionally, let $\bd$ and $\blambda$ be the vectors of diagonal elements of the matrices $\bD$ and $\bLambda$ defined previously. In 
 Metropolis steps, we efficiently evaluate $\log|\tau^2 \bQ(\rho)|$ as 
\begin{align*}
\log|\tau^2\bQ(\rho)| & = - \log |1 / \tau^2\bQ(\rho)^{-1}| \\ 
& = - J \log (1/\tau^2) - \sum \log(\bd \odot (1 - \rho \blambda)),
\end{align*}
where $\odot$ denotes element-wise multiplication. 

\clearpage
\bibliographystyle{apalike}
\bibliography{literature}

\end{document}